\documentclass{aa}

\usepackage{multirow}
\usepackage{newtxtext,newtxmath}
\usepackage[T1]{fontenc}

\DeclareRobustCommand{\VAN}[3]{#2}
\let\VANthebibliography\thebibliography
\def\thebibliography{\DeclareRobustCommand{\VAN}[3]{##3}\VANthebibliography}

\usepackage{graphicx}	% Including figure files
\usepackage{amsmath}	% Advanced maths commands
\usepackage{subcaption}
\usepackage{url}
\usepackage{comment}
\usepackage{hyperref}

\begin{document}
\newcommand{\tdyn}{t_{\rm dyn}}
\newcommand{\Mhalo}{M_{\rm h}}
\newcommand{\aveMhalo}{\bar{M}_{\rm h}}
\newcommand{\nlea}{\bar{n}_{\rm LAE}}
\newcommand{\zpeak}{z_{\rm peak}}
\newcommand{\taure}{\tau_{\rm reion}}
\newcommand{\tauHII}{\tau_{\rm HII}}
\newcommand{\Dss}{\Delta_{\rm ss}}
\newcommand{\Dmod}{\Delta_{\rm mod}}
\newcommand{\Dextr}{\Delta_{\rm extr}}
\newcommand{\GammaHII}{\Gamma_{\rm HII}}
\newcommand{\aveGammaHII}{\langle \Gamma_{12} \rangle_{\rm HII}}
\newcommand{\avenfHII}{\langle x_{\rm HI} \rangle_{\rm HII}}
\newcommand{\fesc}{f_{\rm esc}}
\newcommand{\QHII}{Q_{\rm HII}}
\newcommand{\qnamesevenone}{ULAS J1120+0641}
\newcommand{\sense}{\textsc{\small 21cmsense}}
\newcommand{\dexm}{\textsc{\small DexM}}
\newcommand{\enzo}{\textsc{\small ENZO}}
\newcommand{\Zsun}{Z_\odot}
\newcommand{\uunit}{{\bf \hat{u}}}
\newcommand{\cmfast}{{\tt 21cmFAST}}
\newcommand{\denoiser}{{\tt 21cmPSDenoiser}}
\newcommand{\cmmc}{{\tt 21cmMC}}
\newcommand{\cmemu}{{\tt 21cmEMU}}
\newcommand{\muKK}{\mu {\rm K^2}}
\newcommand{\PkSZ}{[\Delta^{\rm patchy}_{l3000}]^2}
\newcommand{\POV}{[\Delta^{\rm OV}_{l3000}]^2}
\newcommand{\Ptot}{[\Delta_{l3000}]^2}
\newcommand{\reionparams}{\{\zeta, T_{\rm vir}, R_{\rm mfp}\}}
\newcommand{\delz}{\Delta z_{\rm re}}
\newcommand{\zre}{z_{\rm re}}
\newcommand{\HI}{\ion{H}{1}}
\newcommand{\HII}{\ion{H}{2}}
\newcommand{\MgII}{\ion{Mg}{2}}
\newcommand{\OVI}{\ion{O}{6}}
\newcommand{\ef}{x_{e^-}}
\newcommand{\frec}{f_{\rm rec}}
\newcommand{\Ndot}{\dot{N}_{\rm X}}
\newcommand{\Omb}{\Omega_{\rm b}}
\newcommand{\xz}{({\bf x}, z)}
\newcommand{\xzp}{({\bf x}, z')}
\newcommand{\Ts}{T_{\rm S}}
\newcommand{\Tk}{T_{\rm K}}
\newcommand{\aveTs}{\overline{T}_{\rm S}}
\newcommand{\aveTk}{\overline{T}_{\rm K}}
\newcommand{\aveTb}{\overline{T}_{\rm b}}
\newcommand{\denps}{\Delta^2_{\delta \delta}({\bf k}, z)}
\newcommand{\chisq}{\chi^2}
\newcommand{\htwo}{\mathrm{H}_2}
\newcommand{\hone}{\mathrm{H}\textsc{i}}
\newcommand{\hii}{\mathrm{H}\textsc{ii}}
\newcommand{\nf}{\hone}
\newcommand{\Rom}[1]{\uppercase\expandafter{\romannumeral #1}}
\newcommand{\rom}[1]{\lowercase\expandafter{\romannumeral #1}}
\newcommand{\Muv}{M_{\rm 1500}}
\newcommand{\Mvir}{M_{\rm vir}}
\newcommand{\avenf}{\overline{x}_{\hone}}
\newcommand{\nfIGM}{x_{\hone}^{\rm IGM}}
\newcommand{\strom}{HII region}
\newcommand{\lya}{Ly$\alpha$}
\newcommand{\lyb}{Ly$\beta$}
\newcommand{\taudamp}{\tau_{D}}
\newcommand{\taures}{\tau_{R}}
\newcommand{\lobs}{\lambda_{\rm obs}}
\newcommand{\tauobs}{\tau_{\rm obs}}
\newcommand{\tausim}{\tau_{\rm sim}}
\newcommand{\zsource}{z_{\rm s}}
\newcommand{\lcdm}{$\Lambda$CDM}
\newcommand{\fion}{f_{\rm ion}}
\newcommand{\hs}{\hspace{1mm}}
\newcommand{\qnamesfourtwo}{SDSS J1148+5251}
\newcommand{\qnamestwoeight}{SDSS J1030+0524}
\newcommand{\qnamestwotwo}{SDSS J1623+3112}
\newcommand{\qnamestwo}{SDSS J1048+4637}
\newcommand{\qnamefourtwo}{J1148+5251}
\newcommand{\qnametwoeight}{J1030+0524}
\newcommand{\qnametwotwo}{J1623+3112}
\newcommand{\qnametwo}{J1048+4637}
\newcommand{\taub}{\tau_{\rm lim(Ly\beta)}}
\newcommand{\taua}{\tau_{\rm lim(Ly\alpha)}}
\newcommand{\tautot}{\tau_{\rm Ly\alpha}} 
\newcommand{\fobs}{f_\nu^{obs}}
\newcommand{\fcon}{f_\nu^{con}}
\newcommand{\fgamma}{f_\Gamma}
\newcommand{\deltakvec}{\delta({\bf k})}
\newcommand{\deltaxvec}{\delta({\bf x})}
\newcommand{\sigsq}{\sigma^2(M)}
\newcommand{\sig}{\sigma(M)}
\newcommand{\delc}{\delta_c(M,z)}
\newcommand{\Msun}{M_\odot}
\newcommand{\Tvir}{T_{\rm vir}}
\newcommand{\Tvirmin}{T_{\rm vir}^{\rm min}}
\newcommand{\fcoll}{f_{\rm coll}(z, >M_{\rm halo})}
\newcommand{\fcollEPS}{f_{\rm coll}({\bf x}, z, R)}
\newcommand{\Lxsfr}{$\rm{L}_{\rm X}/\rm{SFR}$ }
\newcommand{\deltanlxvec}{\delta_{\rm nl}({\bf x_1}, z)}
\newcommand{\Tcmb}{T_\gamma}
\newcommand{\avedelT}{\delta \overline{T}_b}
\newcommand{\delT}{\delta T_b}
\newcommand{\deldelT}{\delta_{\rm 21}}
\newcommand{\delsq}{\Delta^2_{\rm 21}}
\newcommand{\delNL}{\delta_{\rm nl}}
\newcommand{\Mmin}{M_{\rm min}}
\newcommand{\Mwdm}{m_{\rm wdm}}
\newcommand{\mfp}{R_{\rm mfp}}
\newcommand{\zbegin}{z_{\rm begin}}
\newcommand{\zend}{z_{\rm end}}
\newcommand{\Tgamma}{T_{\gamma, \rm res}}
\newcommand{\avexi}{\bar{\xi}_{12}}
\newcommand{\zon}{z_{\rm on}}
\newcommand{\lmfp}{\lambda_{\rm mfp}}
\newcommand{\lXmfp}{\lambda^{\rm mfp}_{\rm X}}
\newcommand{\toteff}{\epsilon_{\rm fid}}
\newcommand{\taue}{\tau_{\rm e}}
\newcommand{\xhi}{\overline{x}_{\rm HI}}
\newcommand{\Msum}{\mathcal{M}({\bf x}, z)}
\newcommand{\flux}{f({\bf x}, z)}
\newcommand{\Jcrit}{J^{\rm crit}_{21}(M, z)}
\newcommand{\kmps}{\rm km~s^{-1}}
\newcommand{\taueff}{\tau_{\rm eff}^{\rm GP}}
\newcommand\lsim{\mathrel{\rlap{\lower4pt\hbox{\hskip1pt$\sim$}}
        \raise1pt\hbox{$<$}}}
\newcommand\gsim{\mathrel{\rlap{\lower4pt\hbox{\hskip1pt$\sim$}}
        \raise1pt\hbox{$>$}}}
\def\myputfigure#1#2#3#4#5%
{\vskip#5pt\makebox[0pt]{\hskip#2in
\includegraphics[width=#3\textwidth]{#1}}\vskip#4pt\hfill}

\newenvironment{packed_enum}{
\begin{enumerate}
  \setlength{\itemsep}{1pt}
  \setlength{\parskip}{0pt}
  \setlength{\parsep}{0pt}
}{\end{enumerate}}

\newenvironment{packed_item}{
\begin{itemize}
  \setlength{\itemsep}{1pt}
  \setlength{\parskip}{0pt}
  \setlength{\parsep}{0pt}
}{\end{itemize}}

\newcommand{\myurl}{\texttt}
\newcommand{\mytilde}{\texttildelow}

%\newcommand\ion[2]{#1$\;${\scshape{#2}}}%                       % ion, i.e., CII = \ion{C}{ii}

%Journal names
%\newcommand{\apj}{ApJ}
%\newcommand{\apjl}{ApJ}
%\newcommand{\apjs}{ApJS}
%\newcommand{\aap}{A\&A}
%\newcommand{\aj}{AJ}
%\newcommand{\mnras}{MNRAS}
%\newcommand{\pasj}{PASJ}
%\newcommand{\physrep}{Physics Reports}
%\newcommand{\prd}{PRD}
%\newcommand{\prl}{PRL}
%\newcommand{\nat}{Nature}
%\newcommand{\araa}{ARAA}
%\newcommand{\jcap}{JCAP}
%\newcommand{\pasa}{PASA}
%\newcommand{\nar}{NAR}

   \title{Sample Variance Denoising in Cylindrical 21-cm Power Spectra}

   \author{D. Breitman\inst{1}\thanks{E-mail: daniela.breitman@sns.it}
          \and
          A. Mesinger \inst{1}
          \and S. G. Murray \inst{1}
          \and A. Acharya \inst{2}
          }

   \institute{$^1$Scuola Normale Superiore (SNS), Piazza dei Cavalieri 7, Pisa, PI, 56125, Italy \\ 
   $^2$Max-Planck-Institut für Astrophysik, Garching 85748,
Germany
   }

   \date{Accepted XXX. Received YYY; in original form ZZZ}

% Abstract of the paper
\abstract{
%Interferometry with the 21-cm line is set to revolutionise our understanding of the first billion years of the Universe. %To interpret this wealth of data, Bayesian inference frameworks are being developed, and have already been applied on preliminary observations of the spherically-averaged 21-cm power spectrum (1D PS). 
%Current interpretation pipelines focus spherically-averaged 21-cm power spectrum (1D PS) on the However, they all assume a Gaussian likelihood for the 1D PS along with the following approximations: (i) compute the 1D PS of this anisotropic signal by averaging over different cylindrical (2D) wave modes than measured by the observations; and (ii) use one realisation as an estimate of the mean PS. The goal of this work is to propose a solution to these two issues that can be realistically applied in an inference context. 
State-of-the-art simulations of reionisation-era 21-cm signal have limited volumes, generally orders of magnitude smaller than observations. Consequently, the Fourier modes in common between simulation and observation have limited overlap, especially in cylindrical (2D) $k$-space that is natural for 21-cm interferometry. 
This makes sample variance (i.e. the deviation of the simulated sample from the population mean due to finite box size) a potential issue when interpreting upcoming 21-cm observations.
Here, we introduce \denoiser, a score-based diffusion model that can be applied to a single, forward-modelled realisation of the 21-cm 2D power spectrum (PS), predicting the corresponding {\it population mean} on-the-fly during Bayesian inference. 
Individual samples of 2D Fourier amplitudes of wave modes relevant to current 21-cm observations can deviate from the mean by over 50\% for 300 cMpc simulations, even when only considering stochasticity due to sampling of Gaussian initial conditions. \denoiser\ reduces this deviation by an order of magnitude, outperforming current state-of-the-art sample variance mitigation techniques like Fixing \& Pairing 
by a factor of few at almost no additional computational cost ($\sim6$s per PS).
Unlike emulators, \denoiser\ is not tied to a particular model or simulator since its input is a (model-agnostic) realisation of the 2D 21-cm PS.  Indeed, we confirm that  \denoiser\ generalises to power spectra produced with a different 21-cm simulator than those on which it was trained.
To quantify the improvement in parameter recovery, we simulate a 21-cm PS detection by the Hydrogen Epoch of Reionization Arrays (HERA) and run different inference pipelines corresponding to commonly-used approximations.  We find that using \denoiser\ in the inference pipeline outperforms other approaches, yielding an unbiased posterior that is 50\% narrower in most inferred parameters.}%  The current approach of mitigating sample variance by effectively assuming isotropy when computing a spherically-averaged power spectrum can bias the inferred parameters by $>10$\%.   
%: (i) a classical inference such as previous state-of-the-art inferences; (ii) an improved inference where we solve issue (i) only; and (iii) where we solve both issues.  We find that the posterior from inference (i) is biased at over 15$\sigma$ over all parameters combined. In inference (ii), as expected, we find that solving issue (i) completely fixes the bias in the posterior, however, it widens the posterior. Finally, we find that inference (iii) yields an unbiased posterior that is on average $\sim$50\% narrower over each parameter.}

% Select between one and six entries from the list of approved keywords.
% Don't make up new ones.
\maketitle
%%%%%%%%%%%%%%%%%%%%%%%%%%%%%%%%%%%%%%%%%%%%%%%%%%

%%%%%%%%%%%%%%%%% BODY OF PAPER %%%%%%%%%%%%%%%%%%

\section{Introduction}
The cosmic dawn (CD) of the first luminous objects and eventual reionisation of the intergalactic medium (IGM) remain among the greatest mysteries in modern cosmology.
% The CD marks the birth of the first galaxies whose X-rays eventually permeate and heat the intergalactic medium (IGM). It is followed by the epoch of reionisation (EoR) during which the heated neutral IGM is progressively ionised, leading to the present-day Universe. 
%Recent years have seen a dramatic increase in observations of the CD and EoR, including the Lyman-$\alpha$ forest (e.g., \citealt{Fan06,Becker07, Becker15,Bosman18,DOdorico23}), damping wings in quasar spectra (e.g., \citealt{Bolton11, Mortlock11, Banados18, Wang20, Yang20}), and large-scale polarization of the cosmic microwave background (CMB; e.g., \citealt{Planck18, Heinrich21, Belsunce21}). 
One of the most informative probes of the CD and epoch of reionisation (EoR) is the 21-cm line of the hyperfine structure of the neutral hydrogen atom. The 21-cm line has unmatched potential, ultimately able to provide us with a 3D map of more than half of our observable Universe, as expected with the upcoming Square Kilometre Array (SKA\footnote{\url{https://www.skao.int/en}}, e.g. \citealt{Mellema13SKA, Koopmans15SKA, Mesinger19}).

Precursors to the SKA telescope, such as the the Murchison Widefield Array (MWA\footnote{\url{https://www.mwatelescope.org/}}, \citealt{Tingay13}), the Hydrogen Epoch of Reionisation Array (HERA\footnote{\url{https://reionization.org/}}, e.g. \citealt{DeBoer17}), LOw Frequency ARray (LOFAR\footnote{\url{http://www.lofar.org/}}, e.g. \citealt{Haarlem13}), and New Extension in Nançay Upgrading loFAR (NENUFAR\footnote{\url{https://nenufar.obs-nancay.fr/en/homepage-en/}}, e.g. \citealt{Zarka12}) are instead focused on a first detection of the 21-cm power spectrum (PS), since as a well-motivated summary statistic of interferometric observations, it has an enhanced signal-to-noise ratio (S/N) compared to 3D maps. 
Robustly interpreting such measurements is only possible with Bayesian inference.  The current approach to Bayesian inference of 21-cm power spectra, however, relies on several approximations, whose validity is poorly understood\footnote{It is worth noting that these approximations persist in a frequentist framework and have the same detrimental effects as in a Bayesian context, but are more difficult to interpret.}.

Typically, one samples astrophysical and cosmological parameters $\tilde{\theta}$ from priors $p(\theta)$, forward-models the 3D non-Gaussian 21-cm signal with a simulator, compresses the simulated signal into a summary statistic (e.g. the spherically-averaged 1D PS), and compares the forward model to the observed summary, which in the case of the 1D PS is a function of redshift $z$ and Fourier scale $k$: $\Delta^2_{21,\: \rm obs}(k, z$). 
Comparison of the forward model to observations is quantified by a likelihood that is approximated to be a Gaussian at each scale and redshift bin:%$ \mathcal{L}\big(\Delta^2_{21,\: obs} \big| \tilde{\theta}\big)$ for the given PS observation:
\begin{equation}\label{eq:ln}
    \begin{gathered}
        \ln \mathcal{L}\big(\Delta^2_{21,\: \rm obs} \big| \tilde{\theta}\big) \propto - [ \Delta^2_{21,\: \rm obs} - \mu(\tilde{\theta})]^T \Sigma^{-1}(\tilde{\theta})[\Delta^2_{21,\: \rm obs} - \mu(\tilde{\theta})] ~,
    \end{gathered}
\end{equation}
where $\mu(k, z | \tilde{\theta})$ and $\Sigma(k, z | \tilde{\theta})$ are the expectation values and the covariance matrix of the 21-cm PS, averaged not just over modes in given $(k, z)$ bins of a {\it single} simulation, but also averaged over many different {\it realisations}, $i$, of the initial conditions (and any other important source of scatter):
%and are computed from Fourier modes in specified $(k, z)$ bins. In principle,  $\mu(\tilde{\theta})$ and $\Sigma(\tilde{\theta})$ should be averaged over any important source of scatter in the summary, at a fixed parameter vector $\tilde{\theta}$.  
%The most obvious and pervasive source of scatter comes from sampling the initial conditions (ICs) of the simulation, and as such one should average over many different realisations, $i$, of the initial conditions: 
e.g. $\mu(k, z | \tilde{\theta}) = \langle\Delta^2_{21, i}(k, z | \tilde{\theta})\rangle_i$.  However, for computational reasons, sample variance of the initial conditions is ignored i.e. $\mu(\tilde{\theta})$ is generally computed from a single realisation, $\mu(k, z | \tilde{\theta}) \approx \Delta^2_{21, i}(k, z | \tilde{\theta})$,  while the covariance is assumed to be diagonal at some fiducial parameter set, $\Sigma(\tilde{\theta}) \approx \sigma^2 (\theta_{\rm fid})$ (for an in-depth analysis of the accuracy of these approximations, see \citealt{Prelogovic23}).
%Note that $\Sigma^2(\theta_{\rm fid})$ is made up of two components: (i) forward-model variance $\sigma^2_{\rm fm}$: accounts for uncertainty and limitations in the model such as sample variance; (ii) observation sensitivity $\sigma^2_{\rm sens}$: accounts for the imperfection of the instrument via thermal noise, and the fact that there is only one Universe to observe via \textit{cosmic variance}. Cosmic variance is a property of the observable Universe that has its own set of initial conditions and is finite. As such, no matter the size of the field of view of the instrument, there is an intrinsic limitation of the number of modes that can be observed at a given Fourier mode set by the Universe, which introduces scatter in the observations\footnote{Sample variance is the cosmic variance analogue for the forward model.}. 
  The above-mentioned steps are then repeated many times in order to map out the parameter posterior via Bayes' theorem.

%As we move towards a 21-cm detection, the inference pipeline described above can result in sizable biases due to the approximations made when evaluating the likelihood (e.g. \citealt{Prelogovic23}). 
One problem with this approach is that the volume of the forward model does not correspond to that probed by observations.  Due to computational restrictions, simulations which resolve relevant scales have much smaller volumes than observed 21-cm fields. For example, the HERA interferometer observes a volume of over 4 cGpc$^3$, over 100 times larger than typical forward model volumes.
%typical forward models.% that are $\sim$ 0.027 cGpc$^3$. 
 This essentially means that we cannot compare like to like when interpreting observations with theory, resulting in two limitations:
\begin{enumerate}[(i)]
    \item \textbf{sample variance} --- forward models may not have enough independent modes to obtain an accurate estimate of the mean 21-cm PS, $\mu(\tilde{\theta})$, on large scales (i.e. small $k$).%; an unlucky sample of initial conditions could therefore significanly bias choice of initial condition seed could. We find that the sample variance in forward models can be much larger than the cosmic variance of the observation;
    \item \textbf{different PS footprint} --- forward models and observations could probe very different wave modes in cylindrical (2D) Fourier space   (e.g. \citealt{Pober14}).
\end{enumerate}
%Issue (i) arises because forward models do not have enough modes to accurately estimate the 21-cm PS at large scales in comparison to the observation due to their small volume.

Issue (i) can be problematic since initial 21-cm PS detections will likely be limited to a handful of large-scale wave modes; therefore, an "unlucky" realisation of the forward model could result in large biases (e.g. \citealt{Zhao22}; see the detailed analysis in \citealt{Prelogovic23}). 
Naively, sample variance can be mitigated 
%for large enough simulations (e.g. \citealt{Kaur20} cites) 
by taking  an ensemble average over many forward models while varying the initial conditions (ICs; e.g. \citealt{Giri23, Acharya23}), or performing an initial exploration to pick a set of ICs which result in forward model that are close to the mean at some fiducial parameter set $\theta_{\rm fid}$ (e.g. \citealt{Prelogovic23}).    Both of these approaches, however, could require hundreds of additional simulations.  The number of simulations required to accurately estimate the power spectrum mean can be reduced %by pairing their initial conditions.  
by choosing correlated ICs (see \citealt{Racz23} for an overview).
For example, fixing and pairing (F\&P; e.g., \citealt{Angulo16, Pontzen16, Giri23,Acharya23}) requires only one additional evaluation of the forward model. Nevertheless, this is non-negligible computational overhead, considering that typical inferences require over  200k evaluations of the likelihood (e.g., \citealt{HERA22}).
Moreover, the residual uncertainty on the 1D PS mean can still be of order tens of percent even when using paired 21-cm simulations (e.g., \citealt{Giri23, Acharya23}).

Issue (ii) can be problematic since observation and theory cannot be compared in the same region of 2D Fourier space (i.e. $k_\perp, k_\parallel$; see Figure \ref{fig:HERA-window}).  This might not seem like an issue, since the 21-cm likelihood has always been evaluated using the spherically-averaged (1D) PS, comparing theory to the observation at the same magnitude $k = \sqrt{k_\perp^2+k_\parallel^2}$.  However, because the ($k_\perp, k_\parallel$) modes contributing to a given $k$-bin are generally very different for the model and the observation (see Figure \ref{fig:HERA-window}), such a comparison would only be unbiased if the signal were isotropic.
Indeed, the cosmic 21-cm PS is {\it not} isotropic due to redshift-space distortions (RSDs) (e.g. \citealt{Bharadwaj04, Barkana06, Mao12, Jensen13, Pober14, Ross21}), as well as the redshift evolution of the signal along the line of sight direction (e.g. \citealt{Mao12, Datta14, Greig18}). RSDs can boost the 21-cm PS in $k$-modes relevant to observations by up to a factor of $\sim$5 in comparison to the spherically-averaged 21-cm PS at moderate to high neutral fractions (e.g., see Figure 7 in \citealt{Jensen13}). Moreover, ignoring redshift evolution in the 21-cm lightcone could bias inferred constraints by $\sim {\rm few} - 10\sigma$ (e.g. \citealt{Greig18}). 
Therefore averaging the observation and the forward model over different ($k_\perp, k_\parallel$) bins in order to compare them at the same $k$ magnitude might result in sizeable biases.

In this work, we introduce \denoiser\footnote{\url{https://github.com/DanielaBreitman/21cmPSDenoiser}}: a model-independent machine-learning-based tool for sample variance mitigation that provides an estimate of the IC-averaged {\it mean} 2D 21-cm power spectrum given a \textit{single} realisation. %Unlike emulators, \denoiser\ is not tied to a particular model or simulator, and can be applied to any cylindrical power spectra. 
We propose an improved inference pipeline where we mitigate sample variance (issue (i)) by applying \denoiser\ on the simulated 2D PS on-the-fly. Unlike emulators, \denoiser\ is not tied to a particular model or simulator since its input is a (model-agnostic) realisation of the 2D 21-cm PS. %\denoiser\ generalises to out of distribution (OOD) data such as power spectra from different physical models as well as different simulators altogether. %Therefore, unlike emulation, \denoiser\ is not chained to the simulator and physical model it was trained on e.g. via input parameters. These attractive features make our network superior to training an emulator for this task.
To mitigate issue (ii), our pipeline applies a cut in cylindrical $k$-space, averaging only over the modes closest to the ones available in the observation after removing the region dominated by foregrounds and systematics (e.g. \citealt{Pober14}). Cutting out the foreground- and systematic-dominated "wedge" from the forward-modelled 2D PS significantly exacerbates sample variance since it  reduces the number of Fourier modes available when spherically averaging. Indeed, although simulation box sizes larger than $\gtrsim$300 cMpc were found to have negligible sample variance in the spherically-averaged PS (e.g., \citealt{Iliev06,Kaur20}), here we show that is no longer the case after first applying a "wedge" cut in 2D $k$-space.

This paper is organised as follows. We begin by reviewing the traditional inference pipeline in Section \ref{sec:trad_pipeline}. Then in Section \ref{sec:denoiser}, we introduce \denoiser, demonstrating and testing its performance. In Section \ref{sec:fnp}, we compare \denoiser\ to Fixing \& Pairing, a state-of-the-art sample variance mitigation method.
In Section \ref{sec:ood}, we test \denoiser\ on "out-of-distribution" (OOD) data, using power spectra from a different simulator than that used for training. In Section \ref{sec:inf}, we apply \denoiser\ in a realistic inference and compare it to the traditional pipeline.  We conclude with Section \ref{sec:conclusion}, where we summarise the main achievements of this paper. Throughout this work, we assume a flat $\Lambda \rm{CDM}$ cosmology with $(\Omega_\Lambda, \Omega_m, \Omega_b, h,\sigma_8, n_s) = (0.69, 0.31, 0.049, 0.68, 0.82, 0.97)$ consistent with \citealt{Planck18}. All distances are in comoving units unless explicitly stated otherwise.

 \begin{figure*}
\centering
\includegraphics[width=0.95\textwidth]{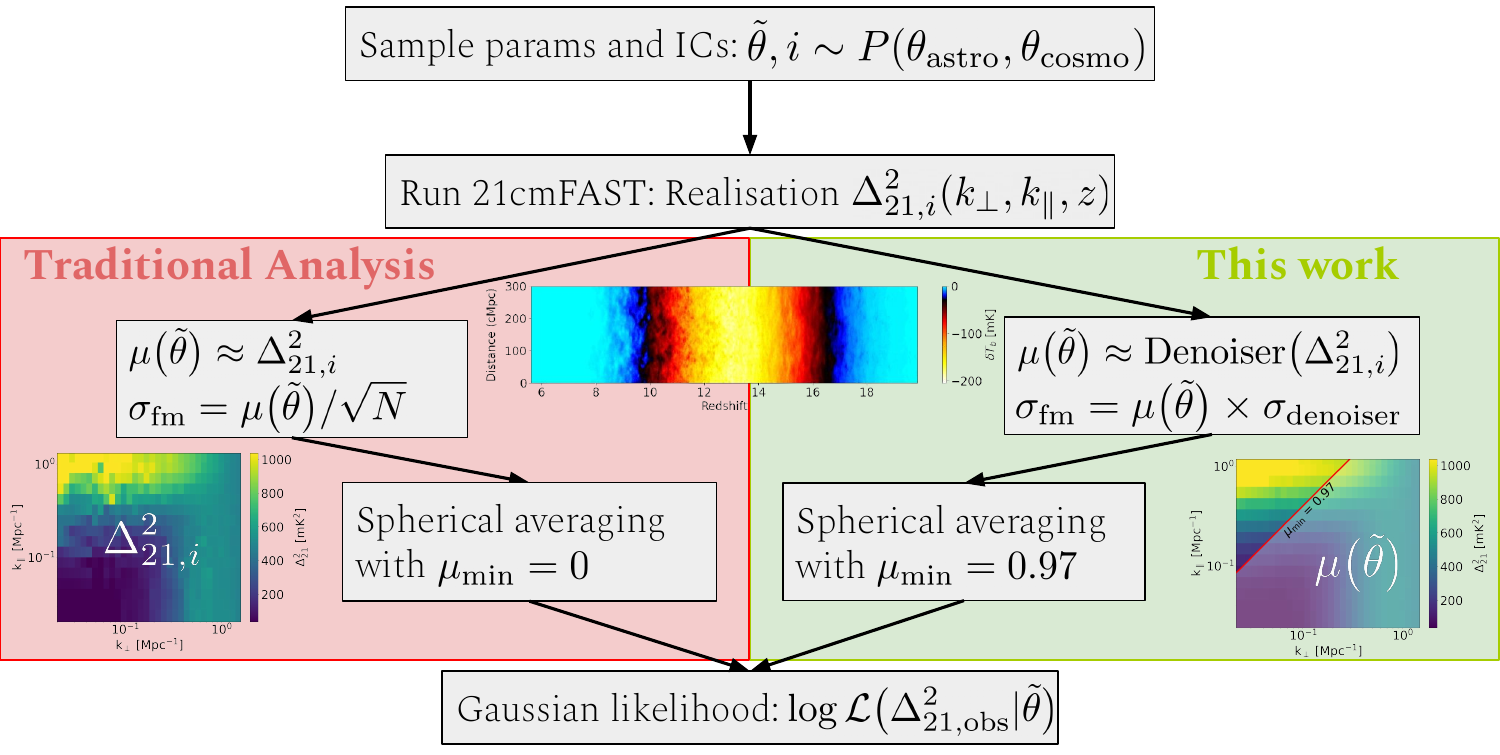}
\caption{Flowchart comparing the current state-of-the-art inference pipeline (left side) with this work (right side). In current state-of-the-art pipelines, we use a single realisation of the initial conditions to estimate the mean 1D 21-cm PS.  Moreover, the simulated 1D PS is computed by spherically averaging over different wave modes than those used to compute the observed 1D PS.
In this work, we account for sample variance by applying \denoiser\, a score-based diffusion model trained to estimate the mean 21-cm PS from a single realisation. We also account for 21-cm PS anisotropy by averaging the 2D PS only over modes above $\mu_{\rm min} = 0.97$, the region of 2D PS that is closest to where current 21-cm PS instruments observe.  Applying a cut in $\mu_{\rm min}$ significantly exacerbates the problem of sample variance, and would not be practical without the use of \denoiser.}
\label{fig:diagram}
\end{figure*}

%%%%%%%%%%%%%%%%%%%%%%%%%%%%%%%%%%%%  SECTION %%%%%%%%%%%%%%%%%%%%%%%%%%%%%%%%%%%%%%%%%%
\section{Explicit likelihood inference from 21-cm power spectra}
\label{sec:trad_pipeline}
In this section, we describe current, state-of-the-art Bayesian inference pipelines (see left side of Figure \ref{fig:diagram}), analogous to those applied to first generation 21-cm interferometers (e.g., \citealt{HERA23, Munshi24, Mertens25, Nunhokee25}). We then introduce the changes proposed in this work (right side of Figure \ref{fig:diagram}).

%Following left side of Figure \ref{fig:diagram}, the first step after sampling a parameter set $\tilde{\theta}$ from the prior $p(\theta)$, is to run the simulator. 

\subsection{Forward modelling the 21-cm signal}
\label{sec:forwardmodel}
A single forward model of the cosmic 21-cm signal consists of the following steps:
\begin{enumerate}
\item Sample astrophysical and cosmological parameters.  For a given set of cosmological parameters, generate initial conditions (e.g. a Gaussian random field sampled from a power spectrum) creating a {\it realisation} of the linear matter density and velocity fields.
\item Evolve densities and velocities to lower redshifts (e.g. via second-order Lagrangian perturbation theory; 2LPT \citealt{Scoccimarro98}). 
\item Assign galaxy properties to the dark matter halos according to the sampled astrophysical parameters (e.g. via semi-empirical relations; \citealt{Park19}). 
\item Compute corresponding inhomogeneous radiation fields and their role in heating and ionising the IGM.
\item Calculate the corresponding 21-cm brightness temperature at each cell and redshift, $(\mathbf{x}, z)$ (e.g, \citealt{Madau97, Furlanetto06, Pritchard12}):
\begin{align}
\label{eq:Tb}
    T_{\rm b}(\mathbf{x}, z) &=  \frac{T_{\rm S} - T_{\rm R}}{1+z}(1-e^{\tau_{21}})\\ 
\nonumber &\approx  27 \ x_{\rm HI} (1 + \delta_b) \left( \frac{\Omega_b h^2}{0.023}\right) \left( \frac{0.15}{\Omega_m h^2} \frac{1+z}{10}\right)^{1/2} \ \rm{mK} \\
\nonumber &\times \left( \frac{T_{\rm S} - T_{\rm R}}{T_{\rm S}}\right) \left[ \frac{\partial_r v_r} {(1+z)H(z)}\right],
\end{align}
where $\tau_{21}$ is the 21-cm optical depth of the intervening gas, and $T_{\rm S}$ and $T_{\rm R}$ are the spin and background temperatures, respectively\footnote{Motivated by observations of local, radio-loud galaxies (e.g. \citealt{Furlanetto06a}) as well as the global 21-cm signal (e.g. \citealt{Cang24, Singh22}), we assume that the radio background is determined by the cosmic microwave background (CMB), therefore $T_{\rm R} = T_{\rm CMB}$.}. $x_{\rm HI}$ is the fraction of neutral hydrogen, $\delta_b \equiv \rho/\bar{\rho} - 1$ is the baryon overdensity, $\partial_r v_r$ is the baryon peculiar velocity gradient along the line of sight,  $H(z)$ is the Hubble parameter at redshift $z$, and $\Omega_m$ and $\Omega_b$ are the mass densities of cold dark matter (CDM) and baryons, respectively. %We note that the brightness temperature is usually calculated in each cell ${\bf x}$ with the equation on the first line, while the second line is a Taylor expansion in the limit of $\tau_{21} \ll 1$ that provides physical intuition.

\item Compute a summary statistic from the 21-cm brightness temperature light cone, in order to compare it to the same summary of the observation.  Currently all\footnote{While there have been studies using the 21-cm 2D PS (e.g. \citealt{Greig24}) and many other summaries (for example the bispectrum e.g., \citealt{Mondal21a,Watkinson22, Tiwari22}, wavelet-based methods e.g., \citealt{Greig22,Zhao24}), and "optimal" learned summaries e.g. \citealt{Prelogovic24,Schosser25}), none of them have yet been applied to real observational data.  Using higher-dimensional summaries would decrease the S/N available for instruments seeking a preliminary detection, as well as making it more important to account for covariances/non-Gaussianity in the likelihood.} analyses use the spherically-averaged 1D 21-cm PS as a summary:
\begin{equation}
    \big \langle \tilde{T}_b (k, z) \tilde{T}^\ast_b (k', z) \big \rangle \equiv (2\pi)^3\delta_D(k - k') \frac{2\pi^2}{k^3} \Delta^2_{21} (k,z)
\end{equation}
where $\tilde{T}^\ast_b (k', z)$ is the conjugate of the Fourier transform of the 21-cm brightness temperature at redshift $z$ and wave mode $k' [\rm{Mpc}^{-1}] = \sqrt{k_x^2+k_y^2+k_z^2} = \sqrt{k_\perp^2+k_\parallel^2}$ for a spherical average and cylindrical average at sky-plane mode $k_\perp$ and line-of-sight mode $k_\parallel$, respectively. $\delta_D$ is the Dirac delta function, and $\Delta^2_{21} (k,z)$ in $\rm{mK}^2$ is the dimensionless 21-cm PS per logarithmic wave mode interval.

\end{enumerate}
Once the forward model is compressed into a summary statistic such as the 21-cm power spectrum, it is compared to the observation by evaluating the likelihood.

\subsection{Evaluating the likelihood on the 21-cm PS}
\label{sec:likelihood}

%In Bayesian inference, the comparison in the last step is done through a likelihood. 
For complex physical models such as the 21-cm signal, the likelihood function is analytically intractable. As such, all current inferences assume a Gaussian functional form\footnote{Simulation-based inference (SBI) can be used to avoid having to define an explicit functional form for the likelihood. However, a Gaussian likelihood for the 1D PS is a decent approximation for preliminary, low S/N data obtained by averaging over many modes and upper limits (e.g., \citealt{Prelogovic23, Meriot24}) such as the data currently available. SBI will be more relevant for high S/N power spectra, as well as for more complicated summaries.} as defined in Eq. \ref{eq:ln}.
Assuming a diagonal variance and approximating the mean 21-cm PS with a single realisation $\Delta^2_{21,\: i}$ can produce a large bias in the resulting posterior for high S/N data (e.g., over 5$\sigma$ as seen from Figure 5 in \citealt{Prelogovic23} ).%\footnote{Even for high S/N power spectra, the Gaussian likelihood can be a very good approximation to the true 1D PS likelihood when the full covariance matrix is evaluated and the simulation seed is chosen such that the 1D PS realisation is close to the mean 1D PS (\citealt{Prelogovic23}).}. Nevertheless, a Gaussian likelihood with a diagonal covariance can be a good approximation to the true likelihood for low S/N observations (e.g. \citealt{Prelogovic23, Meriot24}) such as the ones that we will have over the next few years.

%The use of the Gaussian likelihood is often wrongly motivated by the Central Limit Theorem (CLT) since the averaging on the 21-cm PS is performed on correlated, rather than uncorrelated, modes.\footnote{Recent works (e.g., \citealt{Prelogovic23}) found that the Gaussian likelihood is a very good approximation to the true likelihood when applied with no additional assumptions (e.g. see Fig. 5 in \citealt{Prelogovic23}),}

% The assumptions on the 21-cm PS likelihood were not a big concern in the past since all available observational data of the 21-cm PS were upper limits (e.g. \citealt{Paciga13,Gehlot19, Yoshiura21}) and thus not very constraining. However, with the advent of increasingly better data (e.g. \citealt{HERA22, HERA23}) and a detection on the horizon (e.g. \citealt{Breitman24}), these assumptions are expected to bias the resulting posterior.

\begin{figure}
    \centering
    \includegraphics[width=0.5\textwidth]{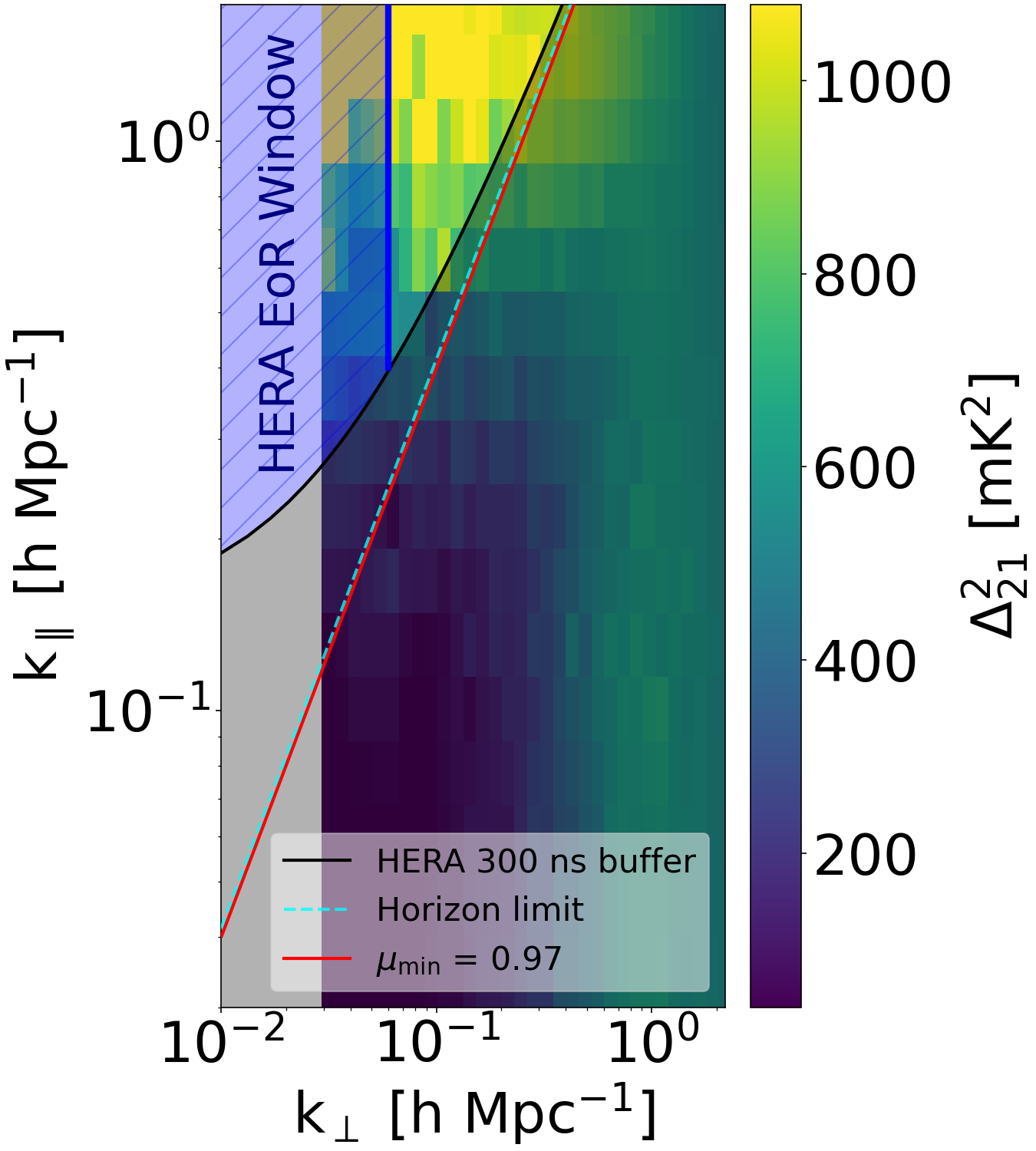}
    \caption{Cylindrically-averaged (2D) 21-cm power spectrum as a function of line-of-sight modes $k_\parallel$ and sky-plane modes $k_\perp$. The color map shows the PS amplitude calculated from a slice through a single simulated light cone centred at $z=9$.  The simulation box has a side length of 300 cMpc and was generated with \texttt{21cmFAST}v3. 
    %The grey shaded area is foreground dominated. 
    The blue hashed area is the HERA EoR window. The dashed cyan line is the horizon limit and the black solid line is the horizon limit with a 300 ns buffer added to it to account for additional foreground leakage (see \citealt{HERA23}). The red solid line is drawn at a value of $\mu_{\rm min} = 0.97$ where $\mu_{\rm min} = \cos \theta$, and $\tan \theta = \frac{k_\perp}{k_\parallel}$. In this paper, we use the red line as a rough approximation for the solid black line.}
    \label{fig:HERA-window}
\end{figure}
%In addition to using a simplified Gaussian likelihood, a
Another common approximation of the traditional analysis is that the 1D PS is computed by averaging the amplitudes of all wave modes in a given $k$-bin.
%over all wavemondes witaveraging  the forward-modelled 21-cm PS over the entire $k$-space available to minimise sample variance when producing the spherically-averaged 1D PS that is compared to the observed PS. 
However, as shown in Figure \ref{fig:HERA-window}, the footprint of 21-cm interferometers in cylindrical (2D) $k$-space can be very different from the corresponding footprint of the forward model.  %available to the forward model is often completely distinct from the $k$-space observed by the instrument (blue shaded region in Figure \ref{fig:HERA-window}). %This is because experiments are designed to tackle the greatest challenge of 21-cm observations - foregrounds. The sky observed by 21-cm experiments includes foregrounds four to five orders of magnitude brighter than the 21-cm cosmological signal itself (cites). Fortunately, these 

The cylindrical $k$-space available to interferometers is limited by a combination of foregrounds, instrument layout, and dish size.  The distribution of baselines limits the accessible angular scales. Foregrounds dominate in the regime of low $k_\parallel$, since they are spectrally smooth.  The chromatic instrument response, however, causes the foregrounds to leak out into a `wedge'-like region in 2D $k$-space (e.g. \citep{Parsons12, Liu14a,Liu14b}). As a result, "clean" EoR measurements are performed outside (or near the boundary) of the wedge (c.f. blue shaded region in Figure \ref{fig:HERA-window}).
%are expected to be spectrally smooth, meaning that they're expected to dominate the power in a wedge-like region of the 2D PS below the so-called horizon limit (black dashed line in Figure \ref{fig:HERA-window}), which defines, for a given sky-plane mode $k_\perp$, the maximum $k_\parallel$ below which a spectrally-smooth signal (such as foregrounds) can contribute to the power. In other words, flat-spectrum emission from the sky cannot contribute to the power above the horizon limit. As such, above the horizon limit, we have the EoR window where we expect the power to be dominated by the cosmological 21-cm signal. 
%There are two main ways to deal with foregrounds: foreground removal methods such as subtraction (cites), and foreground avoidance where we exploit the spectral smoothness of the foregrounds confined to the wedge to exclude them (cites). Current deepest limits on the 21-cm PS are obtained with foreground avoidance (cites). Systematics, imprecise calibration, and poor modelling of the beam frequency response, however, all team up to push foregrounds to leak outside of their wedge and into the EoR window. For example, current HERA observations (e.g. \citealt{HERA23}) add a delay buffer of 300 ns (black solid line, \citealt{Parsons12}) to the horizon limit when defining their EoR window (blue shaded). 
% With improvements to instrument and analysis pipelines, however, upcoming observations will further reduce foreground leakage and edge closer to the horizon limit. 

The cylindrical $k$-space available to forward models is limited by the need to resolve relevant sources and sinks and the corresponding physical processes.  This means that cell sizes of physics-rich simulations cannot be much larger than $\sim$ 1 Mpc.  On the other hand, needing to compute thousands of forward-models in a reasonable time limits the box sizes to $\sim$ few $\times$ 100 Mpc.  Therefore, forward models typically span wave modes of $1 \lesssim \textrm{ k} /{\rm Mpc}^{-1} \lesssim $ few $\times$ 0.01.

This discrepancy between the $k$-space footprint of observations and simulations is problematic because the 21-cm PS is anisotropic due to RSDs and the light cone evolution along the line-of-sight axis. %This anisotropy can boost the averaged 21-cm PS by up to a factor of about 5 in comparison to the 21-cm PS averaged over the same $k$ modes as the observation (e.g. \citealt{Jensen13,Datta14,Pober14,Greig24}). 
This issue can be avoided by cropping the forward-modelled 2D PS to the $k$-space region closest to that of the observation (e.g. the region above the red line in Figure \ref{fig:HERA-window}). However, cropping out modes exacerbates the sample variance problem at large scales as it significantly reduces the number of modes available to perform the averaging. 
%Averaging the 21-cm PS over the entire forward-modelled $k$-space can result in a \textbf{~50\% decrease in power}  (see Figure \ref{fig:1dps}) at scales of $k \sim 0.3 \rm{Mpc}^{-1}$ 

In this work, we introduce \denoiser\ (right branch of Figure \ref{fig:diagram}) to mitigate sample variance: after a forward model is computed, the resulting 2D PS realisation is passed through our neural network that produces an accurate estimate of the mean 21-cm PS. Mitigating sample variance with \denoiser\ allows us to also take into account the 21-cm PS anisotropy: we introduce a cut in cylindrical PS space where we exclude all Fourier modes below $\mu_{\rm min} = 0.97$, marked by a red line in Figure \ref{fig:HERA-window}, where $\mu_{\rm min} = \cos \theta$, and $\tan \theta = \frac{k_\perp}{k_\parallel}$. Introducing this cut significantly reduces the bias due to PS anisotropy (e.g. \citealt{Pober14}).  Restricting the $k$-space footprint comes at the cost of further increasing sample variance and thus further increases the benefits of using \denoiser.

% \begin{figure}
%     \centering
%     \includegraphics[width=0.5\linewidth]{Plots/1dps2.png}
%     \caption{Caption}
%     \label{fig:1dps}
% \end{figure}

%%%%%%%%%%%%%%%%%%%%%%%%%%%%%%%%%%%%  SECTION %%%%%%%%%%%%%%%%%%%%%%%%%%%%%%%%%%%%%%%%%%
\section{Mitigating sample variance with score-based diffusion}
\label{sec:denoiser}

In this section, we first introduce the training database and the neural network architecture of \denoiser. We then evaluate the neural network on a separate database of test samples and comment on its performance.

\subsection{Simulated Dataset}
\label{sec:denoiser_database}
We build a training database with over one hundred light cone realisations (i.e. IC samples) for each set of astrophysical parameters. We simulate a light cone with \cmfast v3\footnote{\url{https://github.com/21cmfast/21cmFAST/}} \citep{Mesinger07, Mesinger11, Murray20} and use ``The Ultimate Eo\textsc{r} Simulation Data AnalYser'' (\texttt{tuesday})\footnote{\url{https://github.com/21cmfast/tuesday}}, a wrapper around \texttt{powerbox}\footnote{\url{https://github.com/steven-murray/powerbox}}\citep{Murray2018}, to calculate the 2D PS.  Our simulation boxes are 300 cMpc on a side, with a cell size of 1.5 cMpc. This is characteristic of the cosmological simulations used to interpret 21-cm observations (e.g. \citealt{Munoz22, Nunhokee25, Ghara25}).

We vary five astrophysical parameters $\theta_{\rm astro} = (f_{\rm esc, 10}, f_{\ast,10}, M_{\rm turn}, L_{\rm X, < 2 keV}/\rm{SFR}, E_0)$, introduced in \citealt{Park19}:
\begin{itemize}
    \item $f_{\rm esc, 10}$: the amplitude of the power-law describing the ionising escape fraction $f_{\rm esc} (M_h) \in [0,1]$  to halo mass relation $f_{\rm esc} (M_h) = f_{\rm esc, 10} (M_h/ M_{10})^{\alpha_{\rm esc}}$, where $M_{10} = 10^{10}M_\odot$ and the index $\alpha_{\rm esc} = 0.5$ is fixed  (e.g., \citealt{Paardekooper15, Kimm17, Lewis20});
    \item $f_{\ast,10}$: the amplitude of the stellar-to-halo mass relation (SHMR) normalised at $M_{10}$.  Similarly to the ionising escape fraction, the faint-end SHMR is described by a power-law whose index we fix to $\alpha_\ast = 0.5$ (e.g. \citealt{Mirocha17, Munshi17, Munshi21});
    \item $M_{\rm turn} [M_\odot]$: the characteristic halo mass scale below which the abundance of galaxies becomes exponentially suppressed to account for inefficient star formation in low-mass halos (e.g., \citealt{Hui97,Springel03, Okamoto08,Sobacchi13,Xu16,Ocvirk20,Ma20});
    \item $\log_{10} \frac{L_{\rm X, < 2 keV}}{\rm{SFR}} \left[\frac{\textrm{erg s}^{-1}}{\rm{M}_\odot \rm{yr}^{-1}}\right]$: the X-ray luminosity escaping the galaxies is modelled as a power-law in energy, which we normalise with the soft-band (i.e. $<2$ keV) X-ray luminosity per unit star formation rate (SFR). We fix the power law index of the X-ray SED to $\alpha_X = 1.0$ (e.g. \citealt{Fragos13,Pacucci14, Das17} );
    \item $E_0 [\rm{eV}]$: minimum energy of X-ray photons capable of escaping their host galaxy;
\end{itemize}
Well-established observations already provide some constraints for $f_{\rm esc}, f_\ast$, and $M_{\rm turn}$. To build our training set, we sample these three parameters from a posterior informed by UV luminosity functions from {\it Hubble} \citep{Bouwens15,Bouwens16,Oesch18}, the Thomson scattering CMB optical depth from {\it Planck} \citep{Planck18, Qin20}, and the Lyman forest dark fraction \citep{McGreer15}.  The resulting posterior is shown in green in Figure 8 of \citealt{Breitman24}).  For the remaining two parameters we sample a flat prior over $\log_{10} L_{\rm X, < 2 keV}/\rm{SFR} \in [38, 42]$ and $E_0 \in [200,1500]$ (e.g. \citealt{Furlanetto06, HERA23}).

The database used for training and validation consists of roughly 900 parameter combinations with an average of roughly 100 realisations per parameter set.  This amounts to $\sim$90k total light cones.
For each light cone realisation, we calculate the 2D PS on cubic chunks over 40 redshift bins $z \in [5.5, 35]$. We re-bin the 2D PS to be linearly-spaced in log scale in both sky-plane and line-of-sight modes. To minimise the number of empty bins while keeping the 2D PS dimensions in powers of two that are more convenient for the NN, we choose 32 $k_\perp$ bins and 16 $k_\parallel$ bins. Due to the re-binning of $k_\perp$ into log-spaced bins, we end up with three empty bins: the second, fourth and fifth. We fill the power in those bins by interpolating with {\tt SciPy} \citep{Virtanen20} which can produce artefacts in individual realisations (e.g. see horizontal bright yellow stripes in the top left plot of Figure \ref{fig:nn_test}). These artefacts, however, do not affect the mean 21-cm PS as they are just another effect of sample variance and get averaged out. Since the NN is redshift-agnostic, the final database consists of about 3.6M 2D PS. This database is then split into 90\% training set with 10\% left for the validation. The test set is made separately from the training and validation sets and is described in Section \ref{sec:test}.

We pre-process the data by applying min-max normalisation on the log of the 21-cm PS and then rescaling it to $[-1,1]$:
\begin{equation}
    \left(\log_{10}\Delta^2_{21, }\right)_{\rm norm} = 2 \times \frac{\log_{10}\Delta^2_{21} - \min (\log_{10}\Delta^2_{21} )}{\max (\log_{10}\Delta^2_{21} ) - \min (\log_{10}\Delta^2_{21} )} - 1.
\end{equation}
The minimum and maximum are each a scalar obtained over the entire training and validation databases.
 
% [Probably rip] But do we actually need 100 realisations? Try training network on 30, 50, etc. realisations and compare w 100.
% PLOT: NN performance vs number of param combinations in the database and number of realisations per parameter.

\subsection{Denoiser architecture and training}
\label{sec:denoiser_train}
\begin{figure}
    \centering
    \includegraphics[width=\linewidth]{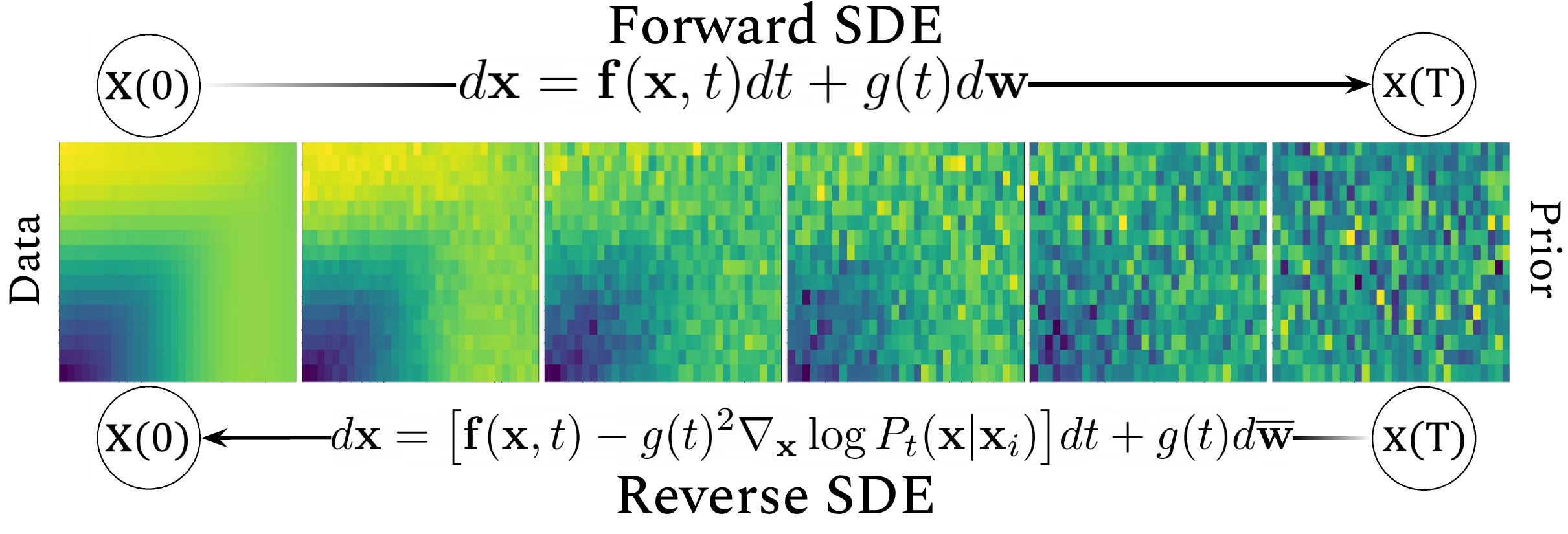}
    \caption{An illustration of the forward and backward diffusion processes used in {\tt 21cmPSDenoiser}. The forward process adds noise to a {\it mean} 21-cm 2D PS sampled from the data distribution ({\it leftmost panel}), transforming it into a pre-defined Gaussian prior distribution ({\it rightmost panel}). We can then write the reverse process that allows us to sample the Gaussian prior and generate a {\it mean} 2D PS, conditioned on the input {\it realisation} of the 2D PS.}
    \label{fig:diffusion}
\end{figure}

We interpret sample variance as a form of non-uniform, (mildly) non-Gaussian (e.g. \citealt{Mondal15, Shaw19}) "noise" added to the target mean power spectrum.  Obtaining the mean cylindrical PS from a single ("noisy") realisation is thus akin to denoising a 2D image.  This is a common task in image processing, for which machine learning is known to outperform traditional methods (e.g. \citealt{Kawar22}) such as Gaussian filter smoothing or principal component analysis (PCA). The idea is to train a neural network (NN) to find a smooth function that interpolates through fluctuating data points, with minimal loss of intrinsic scatter.  In our usage case, the small scales with negligible sample variance (c.f. top right corner of Figure 2) can serve as anchor to the NN prediction of the mean for the noisy larger scales (c.f. bottom left corner of Figure 2).

% how does score-based diffusion work
Here we adopt a state-of-the-art NN architecture for image generation that has also been shown to excel at image denoising (e.g. \citealt{Kawar22}): a score-based diffusion generative model (e.g. \citealt{Sohl15,Song19, SongErmon19, Song19,Ho20,Song20}). In Figure \ref{fig:diffusion}, we illustrate the general idea behind diffusion models: 
\begin{itemize}
    \item Left to right: the forward diffusion process can be interpreted as a continuous noise-adding stochastic process where we corrupt a mean 21-cm PS from the training set with Gaussian noise with increasing variance until it is transformed into a sample from a standard normal prior distribution. This stochastic process can be written as a solution to a stochastic differential equation (SDE): 
\begin{equation}
    d\mathbf{x} = \mathbf{f}(\mathbf{x},t) dt + g(t) d\mathbf{w},
\end{equation}
where $\mathbf{x} = \big(x(0), \hdots, x(T)\big)$ is the diffused data (i.e. a 2D PS) at a given time in the diffusion process, $t \in [0,T]$, with $x(0)$ denoting a sample from the data distribution of mean 2D PS and $x(T)$ a sample from the Gaussian prior. $\mathbf{w}$ is a Wiener process (aka Brownian motion). The drift function $\mathbf{f}(\mathbf{x},t)$ and the diffusion coefficient $g(t)$ are hyperparameters of our model. We choose the most standard drift function and diffusion coefficient leading to the \textit{variance preserving} (VP) SDE \citep{Song20}, which is the continuous-time limit of the Denoising Diffusion Probabilistic Model (DDPM, \citealt{Ho20}).
    \item Right to left: in order to generate new data samples from Gaussian prior samples, we need to reverse the forward process, which can be done by solving the following SDE (\citealt{Anderson1982}):
\begin{equation}
    d\mathbf{x} = \big[\mathbf{f}(\mathbf{x},t) - g(t)^2 \nabla_{\mathbf{x}} \log P_{t} (\mathbf{x}|\mathbf{x}_i) \big] dt + g(t) d\mathbf{\overline{w}},
\end{equation}
where the only unknown is the \textit{score function} $\nabla_{\mathbf{x}} \log P_{t} (\mathbf{x}|\mathbf{x}_i)$ of the probability density function (PDF) $P$ of the data $x$ (in this case, the 2D PS means) explicitly conditioned on a 2D PS realisation $\mathbf{x}_i$ in addition to the continuous diffusion time index $t$.  At $t=0$, the entire procedure boils down to mapping any input 2D PS realisation to its corresponding mean 2D PS.
\end{itemize}

In order to solve the reverse SDE written above and generate a mean 2D PS from a given 2D PS realisation, we train a neural network to learn the score function. Here, we use a U-Net autoencoder architecture\footnote{The model architecture is based on the {\tt PyTorch} implementation available here: \url{https://github.com/lucidrains/denoising-diffusion-pytorch} that is in turn based on the original implementation from \citealt{Ho20} here: \url{https://github.com/hojonathanho/diffusion}} similar to \citealt{Ho20} implemented with {\tt PyTorch} (e.g. \citealt{pytorch}. U-Nets have been developed for image segmentation as they are efficient in recognising local information in images over a range of scales \citep{Ronneberger15}. 
We train the NN with the continuous-time generalisation of the standard DDPM loss function (see Eq. 7 in \citealt{Song20}) with the Adam optimiser \citep{Kingma17}.
%Starting from samples of a known prior Gaussian noise, the NN learns to generate samples of the 2D PS mean.  This process is conditioned on a single input 2D PS realisation, $\mathbf{x}_i =\Delta^2_{21,i}(z)$, and the diffusial time index $t$ (which controls the level of "noise").
%with varying noise levels, $\tilde{\mathbf{x}}(t)$, from Gaussian noise   generated from Gaussian noise while conditioned on a single input 2D PS realisation $\mathbf{x}_i =\Delta^2_{21,i}(z)$ and diffusion time index $t$, to their corresponding mean with different noise levels $x(t)$. At $t=0$, the entire procedure boils down to mapping any input 2D PS realisation to its corresponding mean 2D PS.

Once the NN is trained to accurately predict the score for varying noise levels indexed by $t$, we can generate new mean 2D PS samples given one 2D PS realisation by solving this reverse-time SDE. Since our goal is to use the score-based diffusion model as part of an inference pipeline, we solve the reverse SDE via the probability-flow ODE method \citep{Song20}.  This algorithm modestly sacrifices accuracy for a significant speed increase. In order to average over the network error, our final estimate of the mean 2D PS corresponds to the median obtained over 200 samples (i.e. draws from the prior) from \denoiser\ for a given input.  For these choices, we  obtain a mean 2D PS estimate from a single realisation in $\sim$6 s on a V100 GPU.% We find that we do not benefit from taking more than 100 samples, and so we choose to take 200 samples to be safe since taking more pulls does not increase the runtime of \denoiser.

\subsection{Performance on the test set}
\label{sec:test}

\begin{figure}
    \centering
    \includegraphics[width=\linewidth]{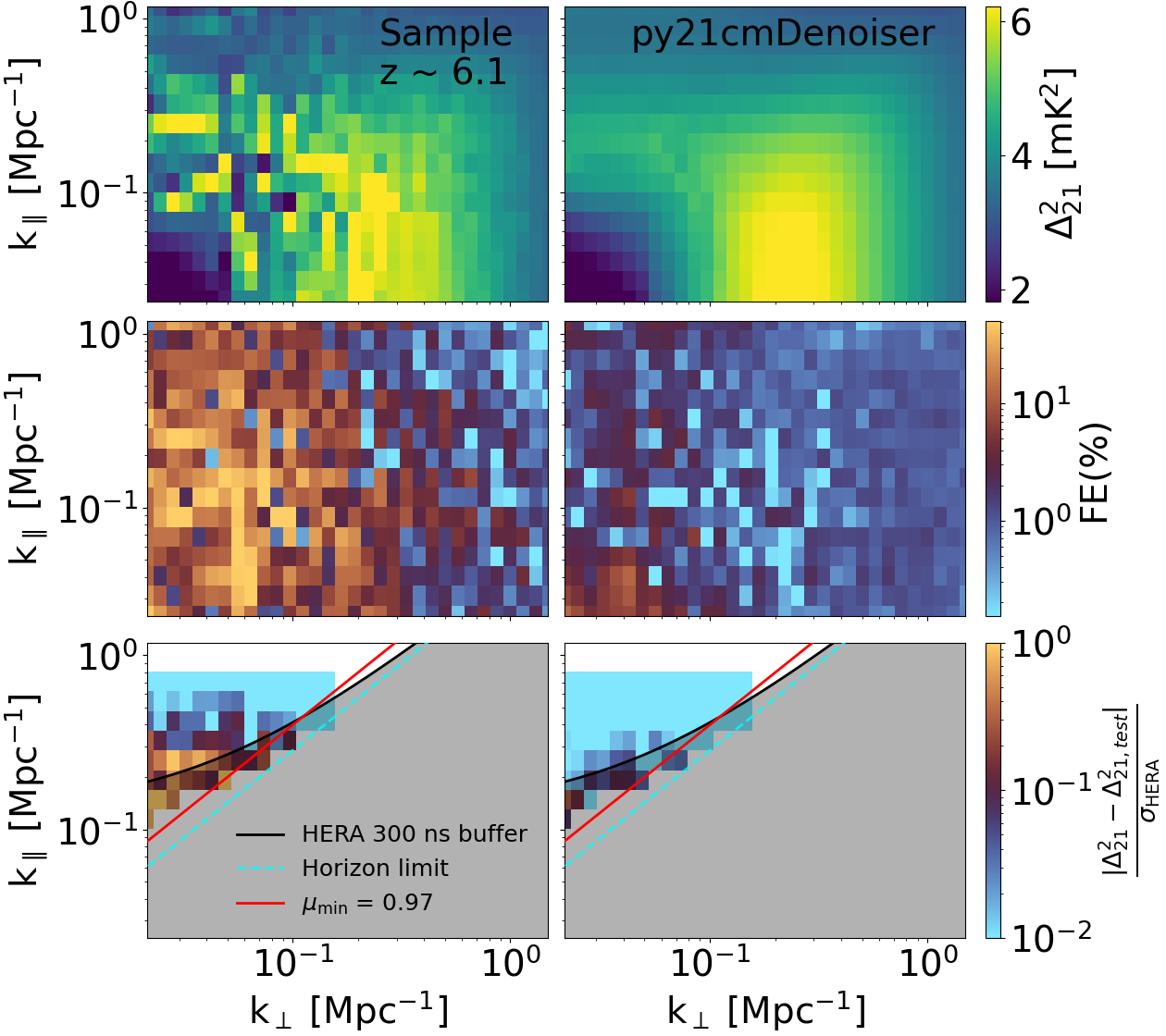}
    \caption{Top row: PS sample on the left and NN mean estimate from this same PS sample as input on the right. Middle row: fractional error with respect to the mean PS obtained from an ensemble average of about 200 PS realisations for the sample (left) and \denoiser\ (right). Bottom row: error as a fraction of the HERA noise level at the same redshift for the sample (left) and for the \denoiser\ (right).}
    \label{fig:nn_test}
\end{figure}

The test set is composed of 50 parameter combinations distinct from the parameters in the training and validation sets with about 200 realisations per parameter and 40 redshift bins for a total of 400k 2D PS.  Since the test set is significantly smaller than the training set, we can afford to have a larger number of realisations for each parameter, allowing a more accurate estimate of the true, target mean PS. %However, since \denoiser\ takes about $\sim 6$ s per 2D PS, we only evaluate it on a subset of the test set, for a total of 40k 2D PS. 
We assess the performance of \denoiser\ using the fractional error (FE) evaluated on a random 40k batch of the test set:
\begin{equation}
    \rm{FE (\%)} = \left|\frac{\Delta^2_{21, test} - \Delta^2_{21, \mu}}{\max 
    \left(0.01, \Delta^2_{21, test}\right)} \right| \times 100  ~ ,
\end{equation}
\noindent where $\Delta^2_{\rm 21, test}$ is the target mean power spectrum, obtained by averaging over 200 realisations per parameter combination, and $\Delta^2_{\rm 21, \mu}$ is a mean 2D PS estimate e.g. from \denoiser. Note that to avoid the fractional error exploding at small power, we floor the denominator to 0.01 mK$^2$, which is an order of magnitude smaller than the accuracy of the \cmfast\ simulator itself (e.g., \citealt{Mesinger11,Zahn11}). We calculate the FE for each realisation from each parameter combination and redshift. %Below, we show the FE distribution from different perspectives, for example, the FE averaged over all test set samples at a fixed redshift bin in Figure \ref{fig:errorz11}, as well as a single example representative of the entire test set in Figure \ref{fig:nn_test}.

In Figure \ref{fig:nn_test}, we illustrate the impact of sample variance using a single PS realisation (left column) and after applying \denoiser\ (right column).  This realisation was chosen from the $\theta_{\rm mock}$ parameter vector in the test set.  This parameter combination is consistent with the most recent constraints from observations such as the Lyman-$\alpha$ forest (e.g.\citealt{Qin25} and UV luminosity functions \citep{Bouwens15, Bouwens16, Oesch18}; see Section \ref{sec:mock} for more details. The top row shows the cylindrical PS realisation at $z\sim6.1$ %\footnote{Note that the PS realisation on the left appears smoother than expected at low $k_\perp$ because the second, fourth and fifth $k_\perp$ bins are actually empty due to the limited size of the simulation box. We fill the power in those bins by interpolating, yielding the horizontal yellow stripes that can be seen on left plot. We do this to maximise the overlap with HERA in cylindrical space (c.f. Figure \ref{fig:HERA-window}). For more details on the interpolation procedure, see the {\tt tuesday} package.} 
and the resulting mean estimate obtained from 
%a median of 200 calls to
\denoiser\ using the realisation shown on the left. In the middle row, we compare them to the test set mean PS via the fractional error where $\Delta^2_{21, \mu} = \Delta^2_{21, \rm NN}$ on the right and $\Delta^2_{21, \mu} = \Delta^2_{21, i}$ on the left. The plotted PS realisation has a median fractional error very close to that of the entire test set for \denoiser\ (see top half of Table \ref{tab:testerr}) and can therefore be considered representative.  We see that the fractional sample variance error on large scales can be reduced by over an order of magnitude by calling \denoiser.

To put these errors into better perspective, the bottom row shows the square root of the squared deviation from the test set mean normalised by realistic HERA sensitivities (see Section \ref{sec:inf} for more details). The left plot on the bottom row shows that the deviation from the mean due to sample variance is comparable to or larger than HERA sensitivity at large scales $k\sim 0.1 \textrm{ Mpc}^{-1}$ and remains significant ($\sim 10 \%$) up to much smaller scales $k \sim 0.5 \textrm{ Mpc}^{-1}$. The right plot shows that applying \denoiser\ reduces sample variance enough that the residual deviation from the mean is very small ($\lesssim 1 \%$) in comparison to the instrument sensitivity.

\begin{figure}
    \centering
    \includegraphics[width=\linewidth]{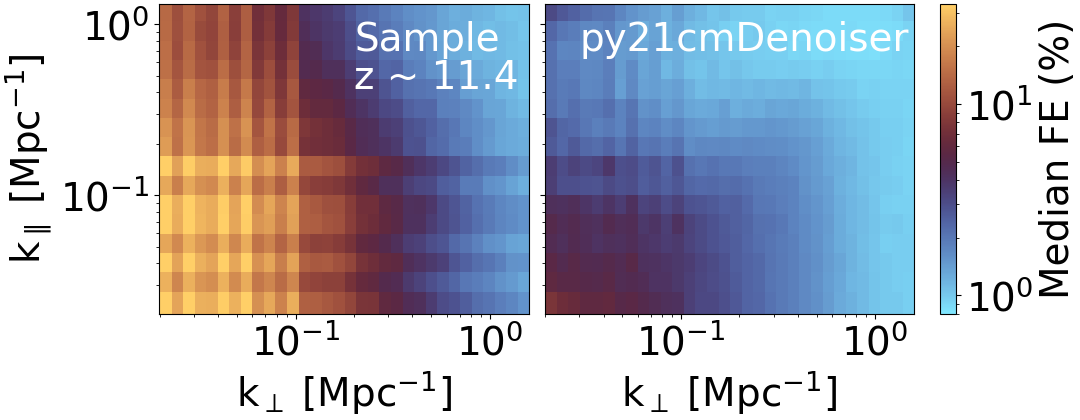}
    \caption{Median fractional error on $\sim 2.5$k test samples at redshift $z \sim 11.4$. The left plot evaluates the FE directly on the PS realisations, while the right plot evaluates it on the output from \denoiser. The striped pattern in the right plot occurs due to the binning scheme, where certain bins have more samples (and thus less sample variance) than others.}
    \label{fig:errorz11}
\end{figure}

In Figure \ref{fig:errorz11}, we show the median of the FE computed from $\sim 2.5$k test samples at redshift $z \sim 11.4$. In the left plot, we evaluate the FE over the PS realisations while in the right plot we evaluate it over the \denoiser\ mean estimate. We can see that when taking the median over $\sim 2.5$k test samples, individual bins can deviate from the mean by over 30\% on large scales and that \denoiser\ reduces this deviation down to $\sim$ 3\%. Individual bins from single PS realisations as shown in Figure \ref{fig:nn_test}, on the other hand, can deviate from the mean by over 50\% on large scales, which gets reduced down to $\sim5$\% by \denoiser. The striped pattern in the right plot occurs due to the binning scheme, where certain bins have more samples (and thus less sample variance) than others.

\section{Comparing \denoiser\ to Fixing \& Pairing}
\label{sec:fnp}
In this section we compare our results against fixing and pairing (F\&P; e.g. \citealt{Angulo16, Pontzen16, Acharya23}), the benchmark technique for mitigating sample variance. F\&P involves \textit{pairing} a simulation to a given \textit{fixed} simulation by reversing the sign of the initial matter overdensity field , $\delta\equiv\rho/\bar{\rho}-1$, such that $\delta_{\rm{paired}}({\bf k}) = - \delta_{\rm{fixed}}({\bf k)}$, for every wave mode $k$.    Each mode whose amplitude is above the mean initial matter PS in one simulation has a counterpart whose amplitude is equally below the mean in the other simulation.  Averaging F\&P simulations by construction would yield the mean initial matter PS for the chosen cosmology, but it has also been shown to give a good estimate of the mean PS of {\it evolved} fields, including galaxy and line intensity maps (e.g. \citealt{Angulo16, Pontzen16, VN18}).

Nevertheless, F\&P has two main shortcomings: (i) it is still computationally expensive, as it doubles the cost of the inference; and (ii) due to non-linear evolution of the 21-cm signal, fixing and pairing becomes less effective with decreasing redshift, which is where current instruments are most sensitive \citep{Acharya23}.

\begin{table}[]

\begin{tabular}{llll}
                 $\mu_{\rm min} = 0$    & Sample & F\&P & NN   \\ \hline
Median FE (\%)       & 7.0     & 4.9   & 2.0    \\
Median AE (mK$^2$)   & 0.3   & 0.2 & 0.1  \\
FE 68 \% CL (\%)     & 26.8     & 21.4      & 8.2      \\
AE 68 \% CL (mK$^2$) & 9.9       & 7.1      & 2.8  \\ \hline
$\mu_{\rm min} = 0.97$    & Sample & F\&P & NN   \\ \hline
Median FE (\%)       & 12.1     & 9.0   & 2.2    \\
Median AE (mK$^2$)   & 0.9   & 0.6 & 0.2  \\
FE 68 \% CL (\%)     & 30.9     & 25.0      & 7.8      \\
AE 68 \% CL (mK$^2$) & 26.2       & 19.4      & 4.6 
\end{tabular}
\caption{Median and 68 \% confidence limit (CL) on the fractional error and absolute error of a $\sim$40k random batch of the test set over 40 redshift bins $\in [5.3, 33]$. The first column shows these quantities when comparing the mean PS directly to the PS realisation. The second and third columns are for the fixing and pairing method and \denoiser\, respectively. The top half of the table ($\mu_{\rm min} = 0$) evaluates the statistics over all cylindrical PS $k$-modes, while the bottom half does so only over the $\mu_{\rm min} = 0.97$ region. }
\label{tab:testerr}
\end{table}

\begin{figure}
    \centering
    \includegraphics[width=\linewidth]{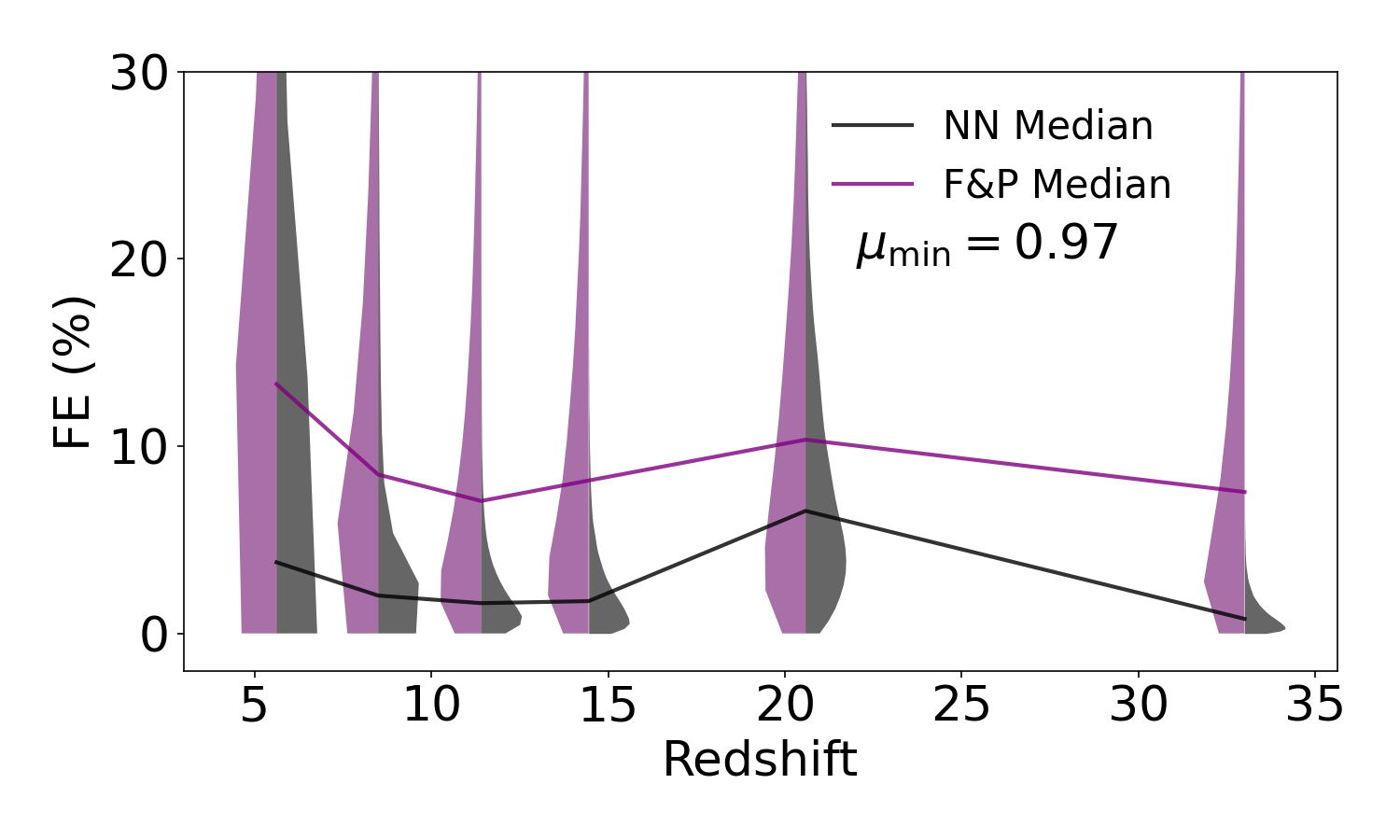}
    \caption{PDFs of the fractional error as a function of redshift for fixing and pairing (purple / left violins) and \denoiser\ (black / right violins) computed on PS modes above $\mu_{\rm min} >0.97$ (see red line in Figure \ref{fig:HERA-window}).  The distributions were generated using the 50 parameter samples comprising our test set.% Note that this comparison only includes power spectra whose mean value, obtained from the entire PS without the $\mu_{\rm min}$ cut, is greater than 0.01 mK$^2$.
    }
    \label{fig:fnpviolin}
\end{figure}

We build a database of 20 F\&P pairs for each of the 50 parameter combinations in the test set. In Figure \ref{fig:fnpviolin}, we compare the fractional error for the fixing and pairing method (purple / left violins) with \denoiser\ (black / right violins). We perform this comparison on power spectra with mean power greater than 0.01 mK$^2$ for $\sim 17$k samples cropped at $\mu_{\rm min} = 0.97$. We can see that \denoiser\ has median fractional error that is about 5\% lower than fixing and pairing across most redshifts\footnote{The increase in FE at redshift $\sim 20$ is due to small values of the 21-cm signal following the dark ages and preceding Lyman-$\alpha$ coupling.}.

We summarise the results in Table \ref{tab:testerr}, where we show the median and 68 \% confidence limits (CLs) of the FE and absolute error (AE) of: (i) individual samples; (ii) the mean estimated from fixing and pairing; and (iii) the mean estimated from \denoiser. The top half of the table shows the FE and AE evaluated over all cylindrical modes, while the bottom half does so over the region above $\mu_{\rm min} = 0.97$.  We see that \denoiser\ results in a FE that is a factor of $\sim 2.5$ ($\sim 4$) smaller than F\&P over all $k$-space ($\mu_{\rm min} = 0.97$ region).  Additionally, unlike F\&P, using \denoiser\ in inference comes at essentially no additional computational cost.

\section{Application to other simulators}\label{sec:ood}

As motivated in the introduction, one benefit of our approach is that, unlike emulation, it is model- and simulator-agnostic.  \denoiser\ can in principle operate on any 2D PS, regardless of what model or simulator was used to make it.  In this section, we test \denoiser\ on hydrodynamic radiative transfer (RT) simulations from \citealt{Acharya23}. Similar to \textsc{thesan} (e.g., \citealt{Garaldi22,Kannan22}), these simulations implement moving mesh hydrodynamics with \textsc{arepo} (e.g., \citealt{Springel10, Weinberger20}), and radiative transfer of ionising photons with \textsc{arepo-rt} (e.g., \citealt{Kannan19}).  Furthermore, as they lack Lyman band and X-ray radiative transfer, these simulations make the simplifying assumption of a  homogeneously-saturated spin temperature: $T_S \gg T_R$ in Eq. \ref{eq:Tb}.   These hydro RT simulations are therefore very different from those used in training \denoiser, both in terms of the source model as well as the simulator.

From \citealt{Acharya23}, we have a total of 40 simulations varying ICs for one parameter set, where five of these also have an additional paired simulation. We post-process the coeval cubes from these simulations into light cones using {\tt tools21cm}\footnote{\url{https://github.com/sambit-giri/tools21cm}} (\citealt{giri20}). These resulting light cones have a resolution of 0.373 cMpc and a box size of 95.5 cMpc. \denoiser, however, is not capable of generalising to a different $k$-space footprint or resolution. As such, we calculate the 2D PS from these light cones with the goal to match the 2D PS from the training database as closely as possible. This involves taking these light cones and first downsampling them by a factor of four so that they have roughly the same resolution as the \cmfast\ simulations in the training set. Next, we calculate the 2D PS on the same $k$-space binning as the training set. However, since these boxes are about three times smaller, this leaves many large-scale mode bins empty. We pad these large-scale bins by copying over the power from the closest non-empty bin at smaller scales. This padded 2D PS is then passed to \denoiser \footnote{The scripts used to make these power spectra use the publicly available \tt{tuesday} package and can be found in the github repo \url{https://github.com/DanielaBreitman/21cmPSDenoiser}}. 

\begin{figure}
    \centering
    \includegraphics[width=\linewidth]{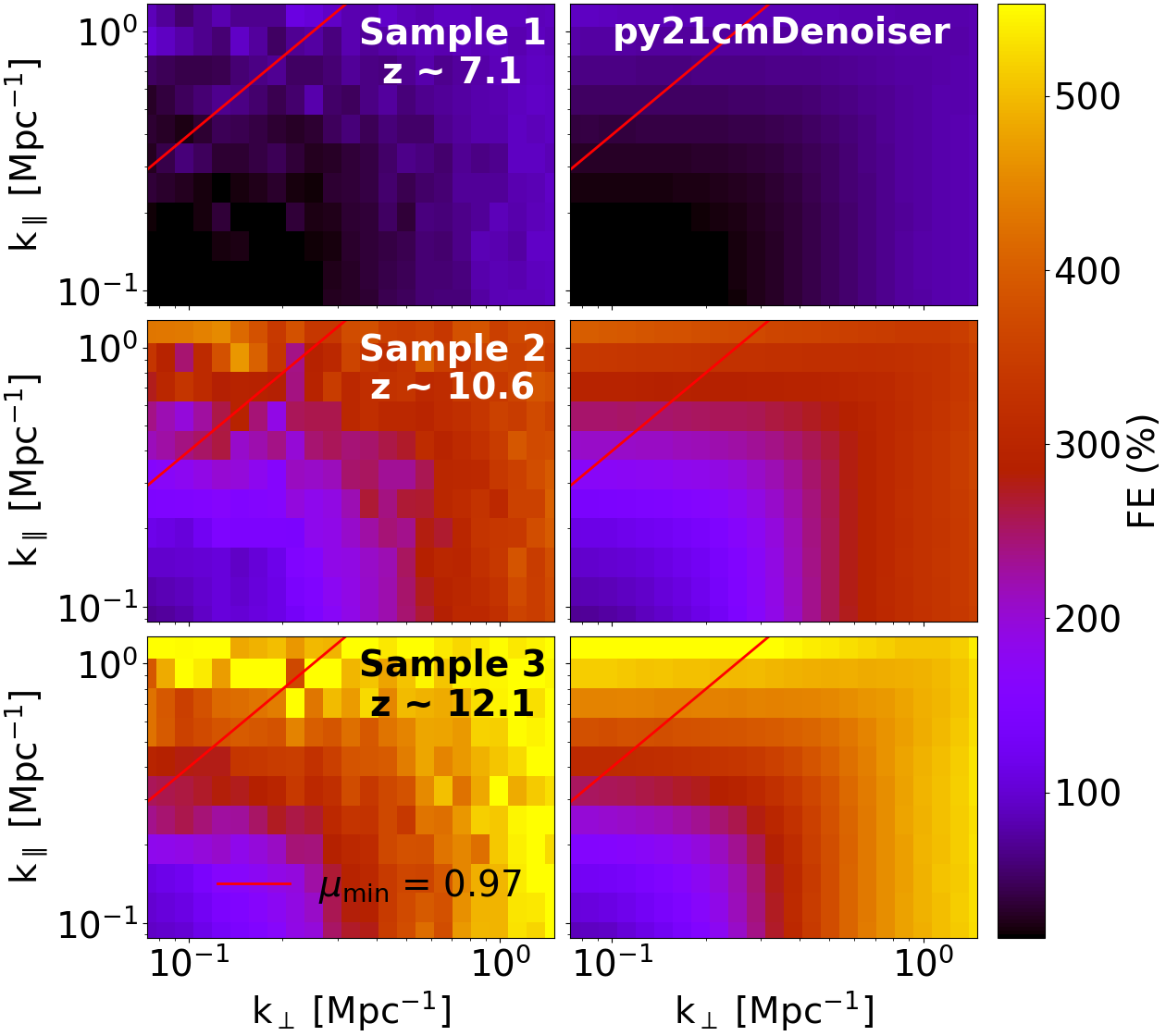}
    \caption{21-cm PS realisations from a hydro RT simulation (left column) and the corresponding mean estimate (right column) obtained after passing the realisation on the left to \denoiser\ (which was only trained on \texttt{21cmFAST}). Each row corresponds to different initial conditions and redshift.}
    \label{fig:ood}
\end{figure}

% We find that \denoiser\ always has a smaller FE than a single realisation when compared to the \textit{true} mean estimated from all 45 simulations over the $\mu_{\rm min} > 0.97$ region of the 2D PS. 
In Figure \ref{fig:ood}, we show the de-noised PS on the right, corresponding to the single hydro RT realisation input shown on the left.  Each row corresponds to a different realisation. 
Note that in all plots, we crop the $k$-space footprint to only include the non-empty bins (i.e. we do not consider the padded bins as we have no larger hydro RT simulations against which to compare them). We note that with only 35 realisations and 5 F\&P pairs, the estimate of the true target mean is not as accurate as it is for the test set, where we have $\sim 200$ realisations per parameter. As a consequence, the performance of the NN reported in this section is not as accurate as in previous sections.
% In the top row, we show one 21-cm PS realisation on the left and the mean estimate obtained from the network on the right. In the middle row, we evaluate the FE of each against the mean obtained from all 45 RMHD simulations available. On the bottom row, we show the deviation from the mean normalised by SKA1-low AA4 noise that we simulate with {\tt 21cmSense} (Breitman et al. (in prep.)). Note that the mean 21-cm PS used to evaluate the FE is a bit noisy as it is obtained from averaging only 35 realisations with the 5 fixing and pairing pairs, while \denoiser\ was trained to produce 2D PS means obtained from over a hundred realisations. It is thus possible that the FE on the NN is somewhat inflated given the smaller number of samples available to estimate the \textit{true} mean 21-cm PS. 

%We see from Figure \ref{fig:ood} that \denoiser\ does a good job in mitigating sample variance even on such 2D PS realizations computed with a different simulator, assuming a different astrophysical model than used for training.  The samples shown in the figure were chosen to span a wide range of input signals, and we confirm that the illustrated performance of  \denoiser\ is representative of the available dataset.

\begin{figure}
    \centering
    \includegraphics[width=\linewidth]{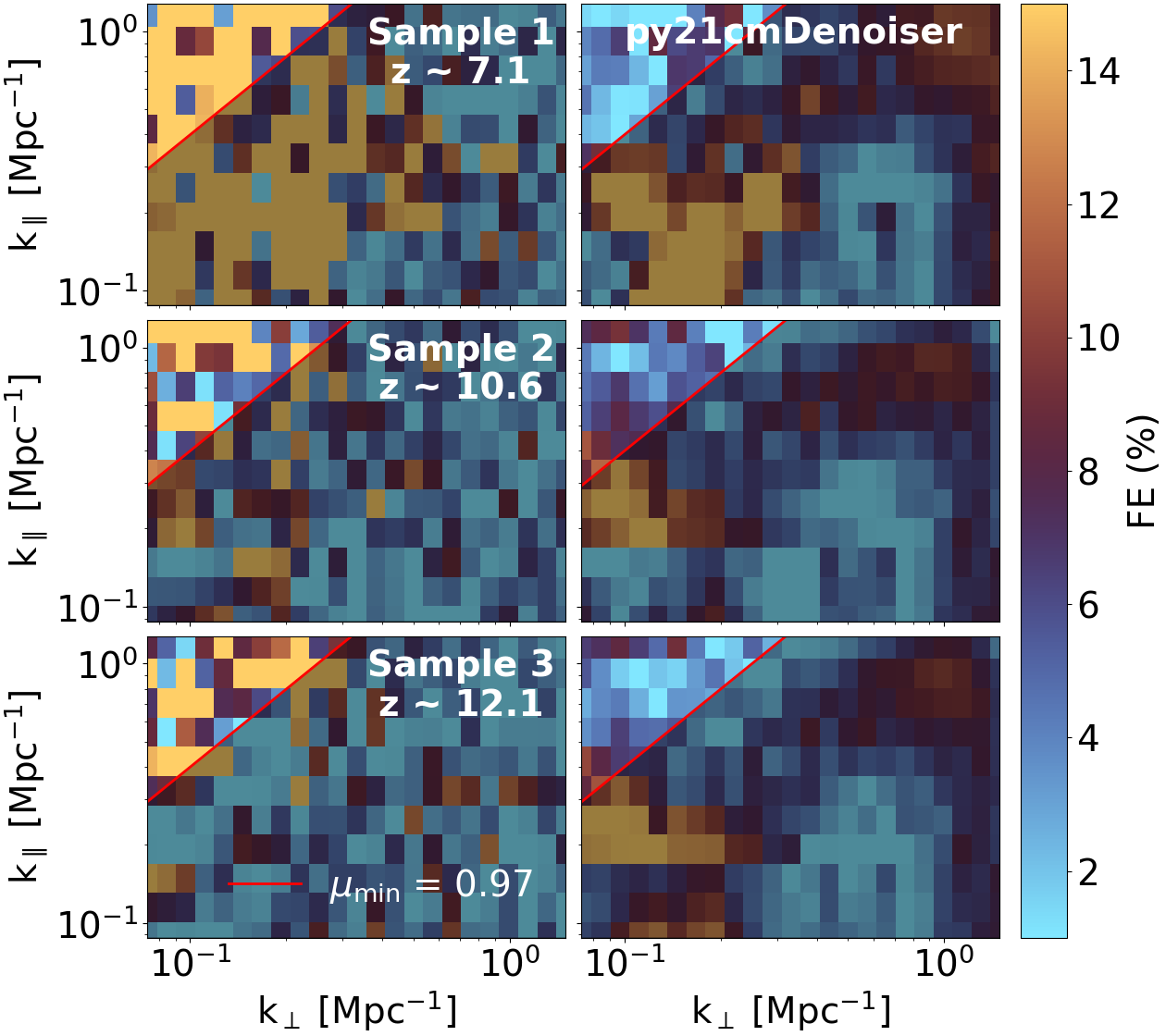}
    \caption{Fractional error on the 21-cm power spectra shown in Figure \ref{fig:ood}. We caution that the mean PS is estimated from a relatively small number of realisations (35 unpaired ICs + 5 pairs of ICs).}
    \label{fig:ood_fe}
\end{figure}

We see from Figure \ref{fig:ood} that \denoiser\ is flexible enough to mitigate sample variance on 21-cm power spectra from different astrophysical models and different simulators. 
We confirm this by looking at Figure \ref{fig:ood_fe}, where we plot the FE of the samples shown in Figure \ref{fig:ood}.  The mean PS above $\mu_{\rm min} = 0.97$ is recovered to an median accuracy of about $\sim4$\% over all redshifts and PS realisations for the denoiser, while sample variance introduces a median deviation of about $\sim 10$\% in the same region.  Note that the samples shown have an average FE at the redshift plotted that is slightly above the average FE over all available realisations. Figure \ref{fig:ood_fe} also shows that on the largest and smallest scales, \denoiser\ tends to under-predict the mean power: a behaviour not seen in the test set cases.   It is reasonable, however, that the out-of-distribution performance is weaker than that seen for the test set.  One could further improve the generalization of \denoiser\ by fine tuning it on 21-cm power spectra from different models and simulators.

% Finally, we would also like to highlight that \denoiser\ can be used to inpaint large-scale modes on smaller simulations such as the ones discussed in this section. Although we had to pad the input 2D PS realisation before passing it to the NN, the output mean 2D PS looked reasonable at all scales. While we cannot test this with the RMHD simulations, we verify this with the test sample from Figure \ref{fig:nn_test}. We find that the NN can inpaint large-scale modes up to $k_\perp \sim 0.05 \textrm{ Mpc}^{-1}$ and $k_\parallel \sim 0.1 \textrm{ Mpc}^{-1}$ at a similar accuracy and precision \textit{as if its input were the entire 2D PS realisation}. In other words, for \texttt{21cmFAST} simulations, we could simulate a 100 cMpc box and use \denoiser\ to obtain a mean 2D PS of a similar quality as if we have simulated a 300 cMpc box.

%We briefly check this using the same sample as shown in Figure \ref{fig:nn_test}. We take the original sample (top left plot of Figure \ref{fig:nn_test}) and replace the first $N$ $k_\perp$ ($k_\parallel$) columns (rows) with the values present at the $(N+1)$th column (row). We find that the NN inpaints the power spectrum at an accuracy and precision similar to what is shown in Figure \ref{fig:nn_test} for up to $N = 5$. At $N > 5$, the performance begins to degrade. Nevertheless, on average, up to $N=6$ the inpainted PS values are still closer to the true mean than the those from the realisation. We can thus conclude that \denoiser\ can be reliably used to inpaint the first five $k_\perp$ columns and $k_\parallel$ rows \textit{for 21cmFAST simulations with similar resolution and astrophysics as the training database}. We leave testing this claim on power spectra from different simulators for future work.

\section{Application to inference}
\label{sec:inf}

In this section, we test the performance of \denoiser\ by performing inference on a HERA mock observation. We compare the traditional state-of-the-art inference pipeline with the improvements introduced in this work (i.e. left vs right of Figure \ref{fig:diagram}).

\subsection{Mock HERA observation}
\label{sec:mock}

We choose the mock parameter set $\theta_{\rm mock}$ with $(\log_{10} f_{\rm esc, 10}, \log_{10} f_{\ast,10}, M_{\rm turn}, \log_{10} L_{\rm X, < 2 keV}/\rm{SFR}, E_0) = (-1.23, -1.36, 8.26, 40.59, 1.40 \textrm{keV})$ out of the 50 parameter sets available in the test set as it has an EoR history that is consistent with that inferred from Lyman-$\alpha$ forest data \citep{Qin25}, and matches UV luminosity functions at $z=6$--10  \citep{Bouwens15, Bouwens16, Oesch18}.  In addition to thermal variance from the instrument, an observation of the 21-cm PS is subject to \textit{cosmic variance} due to the fact that there is only one Universe with its own set of initial conditions to observe i.e. we do not observe the expectation value of the 21-cm PS (over all possible observable universes), but rather a realisation with a finite volume. Cosmic variance has the most effect at the largest scales of an observation as there are fewer Fourier modes to be observed given the finite size of the observable Universe. However, since HERA observations span a much larger field than our forward models, cosmic variance on the scales of the simulation is negligible.  Our mock cosmic signal is therefore taken to be the mean 21-cm PS, averaged over the 200 realisations of ICs for $\theta_{\rm mock}$.    As our observational summary statistic, we spherically-average this mean cylindrical PS above $\mu_{\rm min} = 0.97$, mimicking the observational footprint (c.f. Figure \ref{fig:HERA-window}), obtaining the 1D PS at $z=5.6, 6.1, 6.9, 7.9, 9.1, 10.4, 10.8, 16.8, \textrm{ and }22.7$.

We use {\tt 21cmSense} (\citealt{Pober13, Pober14b, Murray24}) to forecast the sensitivity for two full seasons of phase II HERA observations where we observe for 94 nights per season for a total of $\sim 2256$ integration hours (see \citealt{HERA23} and the appendix in \citealt{Breitman24} for more details). The forecast assumes that the number of observed hours is the same in both seasons. The only difference between the two seasons is the number of operating antennas that increased from 140 in the 2022-2023 season to 180 in the 2023-2024 season. To combine the two observations, we sum the total integration times as well as the $uv$ coverage of both seasons. We then use these combined integration time and $uv$-coverage to obtain the thermal variance of the HERA instrument over both seasons. We also include the cosmic variance of the observation in the error budget, adding it to the thermal noise in quadrature: 
 $\sigma_{\rm sens} = \sqrt{\sigma^2_{\rm thermal} + \sigma^2_{\rm cosmic}}$, where we refer to $\sigma_{\rm sens}$ as the \textit{sensitivity}.

\begin{figure}
    \centering
    \includegraphics[width=\linewidth]{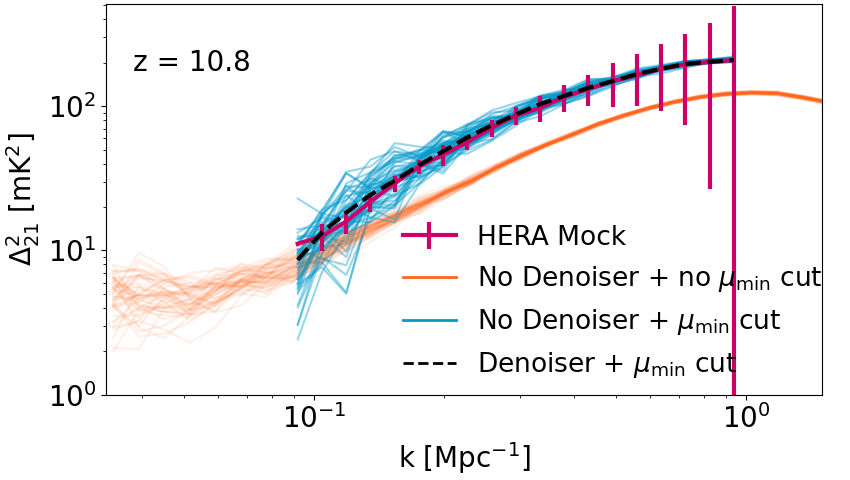}
    \caption{Spherically averaged 21-cm power spectra at $z=10.8$ corresponding to a parameter vector in our test set, $\theta_{\rm mock}$.  Pink points and error bars correspond to a mock $\sim 2256$ h observation with HERA (see text for details). In orange, we plot different realisations, varying ICs at a fixed $\theta_{\rm mock}$, but spherically-averaging the 2D PS down to $\mu_{\rm min} = 0$ as is commonly done when forward-modelling. In blue, we plot these realisations but instead averaging only down to $\mu_{\rm min} = 0.97$  to account for the HERA footprint in cylindrical space (see red line in Figure \ref{fig:HERA-window}).  Excising low $\mu$ modes removes the bias seen in the orange curves, but dramatically increases the sample variance.  The dashed black line corresponds to the output of \denoiser\ from a single 2D PS realisation, averaged down to $\mu_{\rm min} = 0.97$.  We see that applying \denoiser\ mitigates both the bias and the sample variance.}
    \label{fig:hera_mock}
\end{figure}
After this procedure, we are left with HERA mock 1D 21-cm power spectra at all redshifts, together with the associated sensitivities. 
In Figure \ref{fig:hera_mock}, we plot the mock observation at $z=10.8$ as the pink points with 1$\sigma$ error bars. In orange, we show different PS realisations with $\theta = \theta_{\rm mock}$, where all modes (i.e. $\mu_{\rm min} = 0$) of the 2D spectrum have been included in the average, instead of the modes above $\mu_{\rm min}=0.97$ used for the mock.  
Aside from the realisation-to-realisation scatter between the orange curves, we find that they are biased by a factor of $\sim3$, revealing the effects of anisotropy when the correct $\mu_{\rm min}$ is neglected.\footnote{We note that we chose $z=10.8$ explicitly to highlight the impact of PS anisotropy; its impact at other redshifts corresponding to an advanced stage of the EoR and whose bins span a slower redshift evolution of the signal would be smaller (e.g. \citealt{Mesinger11, Mao12, Datta14}).}

In blue, we show PS realisations also with $\theta = \theta_{\rm mock}$, but now averaging over only the modes `observed' in the mock with $\mu_{\rm min} = 0.97$.  
Unlike the orange curves, these do not display a significant bias with respect to the mock data. 
However, limiting the modes over which to perform the averaging results in substantial sample variance at small $k$.  
The level of sample variance is well in excess of the observational error bars and we thus expect it to limit the inferred parameter constraints (as we confirm below).

 The dashed black line shows the mean PS obtained from \denoiser\, derived from a single realisation.  
 Similarly to the blue curves, this de-noised PS is averaged only over modes above $\mu_{\rm min}=0.97$.
% As for the blue curves, we average the \denoiser\ outputted 2D PS down to $\mu_{\rm min} = 0.97$ to obtain the corresponding 1D PS.  
The de-noised PS agrees very well with the mock data, with no obvious bias and a dramatic reduction of the effects of sample variance (seen by the reduction in bin-to-bin variance) compared with the blue curves.
 
 %applying the $\mu_{\rm min} = 0.97$ cut on the mean 2D PS estimate from \denoiser\ (i.e. right branch of Figure \ref{fig:diagram}). We obtain the mean 2D PS estimate from the NN by taking the median over 200 mean estimates for a given PS at a given redshift. This plot shows that 21-cm PS anisotropy can incur biases of over 50\% at scales $k \gtrsim 0.2$ and moderate neutral fractions ($x_{\rm HI} \sim 0.5$ in this case). Moreover, attempting to mitigate 21-cm PS anisotropy without mitigating sample variance can lead to a bias of over 50\% at large scales $k \lesssim 0.2$ Mpc$^{-1}$. Applying \denoiser\ for sample variance mitigation and the $\mu_{\rm min}$ cut to account for 21-cm PS anisotropy yields a much closer match to the mock. 

\subsection{Inference set-up}

To quantify the qualitative trends seen in Figure \ref{fig:hera_mock}, we use our HERA mock observation to perform inferences under different approximations:
\begin{enumerate}[(i)]
    \item \underline{No \denoiser\ and no $\mu_{\rm min}$ cut} -- this corresponds to the current approach of spherically-averaging the forward-modelled 2D PS down to $\mu_{\rm min}=0$ using a single realisation of the ICs (c.f. orange curves in Figure \ref{fig:hera_mock});
    \item \underline{No \denoiser\ with $\mu_{\rm min} = 0.97$ cut} -- this also uses a single realisation of the 2D PS, but averages down to the same $\mu_{\rm min}=0.97$ as is used for the mock data (c.f. blue curves in Figure \ref{fig:hera_mock});
    \item \underline{\denoiser\ with $\mu_{\rm min} = 0.97$ cut} -- this uses a single realisation of the 2D PS passed to \denoiser\ to obtain the mean 2D PS, before averaging down to the correct $\mu_{\rm min} = 0.97$ (c.f. dashed black curve in Figure \ref{fig:hera_mock}).
\end{enumerate}

In all of the inferences, the likelihood is constructed by multiplying individual likelihoods based on four observables:
\begin{enumerate}[(i)]
    \item \textbf{21-cm 1D PS}: following Figure \ref{fig:diagram}, we forward model the 2D PS $\Delta^2_{21,i}(k_\perp,k_\parallel,z)$. We then estimate the mean PS:
    $\mu(\theta) = \begin{cases}
        {\rm Denoiser}(\Delta^2_{21,i}) \textrm{ with NN}\\
        \Delta^2_{21,i} \textrm{ otherwise.}
    \end{cases}$ \\ 
    We then average the mean 2D PS to 1D, weighting each 2D bin by the number of 3D Fourier modes it contains, $N_k$:
    \begin{align}
        \Delta^2_{21} (k_i) = \frac{1}{\sum_{k \in \mathcal{K}_i} N_k(k_\perp,k_\parallel)}\sum_{k \in \mathcal{K}_i} N_k(k_\perp,k_\parallel) \times \Delta^2_{21} (k_\perp, k_\parallel), 
    \end{align}
    where $\mathcal{K}_i$ is the set of $k = \sqrt{k_\perp^2 + k_\parallel^2} \in k_i$\footnote{For more details, see \texttt{cylindrical\_to\_spherical} in \texttt{tuesday}.}. We then apply a window function $W$ (e.g., \citealt{Liu14a, Liu14b, Gorce23}) calculated with \texttt{hera-pspec}\footnote{\url{https://github.com/HERA-Team/hera_pspec/}} to the forward-modelled 1D PS. The window function converts the forward-modelled 21-cm PS to an \textit{observed} PS by including   instrumental effects such as the chromaticity of the baselines and the beam, especially affecting large-scale modes. The likelihood function is a Gaussian
    \begin{equation}
        \log \mathcal{L} (\Delta^2_{21,\rm mock} | \theta) = -\frac{1}{2} \sum_{f,k} \frac{\left( W \cdot \mu(\theta) - W \cdot \Delta^2_{21,\rm mock}\right)^2}{\sigma^2},
    \end{equation}
    summed over all frequency bands $f$ and $k$-bins , where the total variance $\sigma^2$ includes the sensitivity of the mock observation $\sigma^2_{\rm sens}$ and a contribution from the forward model $\sigma^2_{\rm fm}$ coming from either Poisson sample variance or the mean NN error: \begin{align}
        & \sigma^2 = \sigma^2_{\rm sens} +\sigma^2_{\rm fm}, \textrm{ where} \\ 
        & \sigma_{\rm fm} = \begin{cases}
        \Delta^2_{21}(\theta) \times \sigma_{\rm denoiser} \textrm{ with \denoiser}\\
        \Delta^2_{21}(\theta)/\sqrt{N} \textrm{ otherwise.}
    \end{cases} 
    \end{align}  %\\ In other words, when the NN is used, the sample variance is represented by the uncertainty in the NN, and when the NN is not used, the sample variance is estimated via the Poisson error on the 21-cm PS.
    \item \textbf{UV luminosity functions}: we include UV LFs at $z = 6,7,8 \textrm{ and } 10$ based on {\it Hubble} data (\citealt{Bouwens15, Bouwens16, Oesch18}) as in previous works (e.g., \citealt{HERA22, Breitman24}). This likelihood function is also assumed to be Gaussian.
    \item \textbf{Thomson scattering optical depth to the CMB}: we include a Gaussian likelihood centred around $\tau_e = 0.0569^{+0.0081}_{-0.0086}$ based on the median and 68\% credible interval (CI) from the posterior obtained in \citealt{Qin20} from their re-analysis of \citealt{Planck18} data.
    \item \textbf{Lyman forest dark fraction}: this term compares the proposed model's global neutral fraction at $z=5.9$ with the upper bound $\overline{x}_{\textsc{hi}} < 0.06 \pm 0.05$ at 68\% CI obtained with the model-independent QSO dark fraction method (\citealt{McGreer15}). The likelihood function is unity if the proposed global neutral fraction is below the upper bound at $z=5.9$, then it decreases as a one-sided Gaussian for higher values of $\overline{x}_{\textsc{hi}}$. 
\end{enumerate}

\begin{figure}
    \centering
    \includegraphics[width=\linewidth]{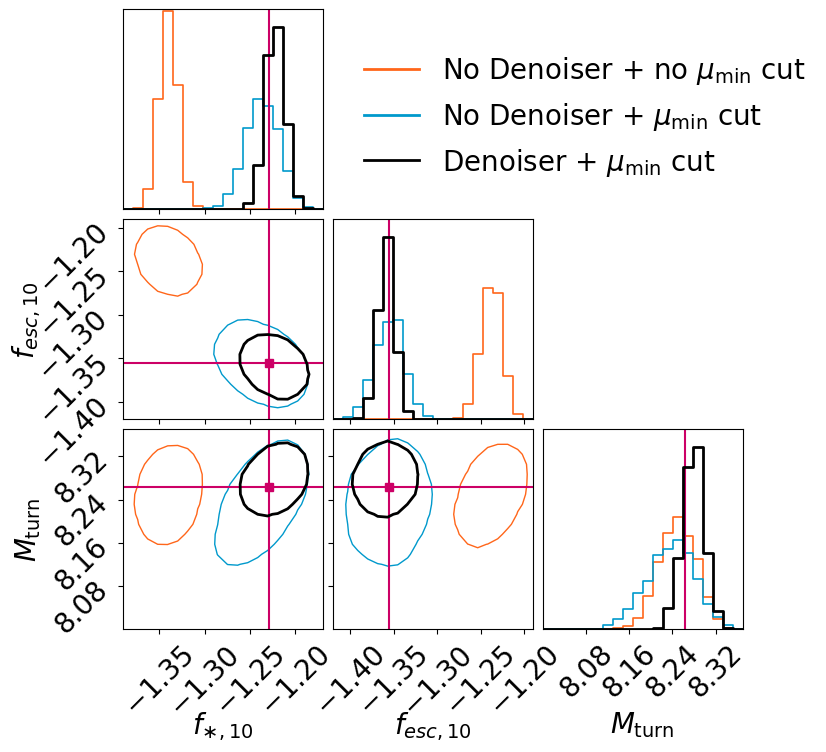}
    \caption{1D and 2D marginal posteriors from three inferences described in the text: (i) traditional state-of-the-art analysis using a single IC realisation and spherically-averaging the 2D PS down to $\mu_{\rm min}=0$ (orange); (ii) using a single IC realisation but spherically-averaging the 2D PS down to the same $\mu_{\rm min}=0.97$ used for the mock data (blue);  and (iii) applying \denoiser\ to a single realisation in order to mitigate sample variance followed by the $\mu_{\rm min}$ cut to mitigate PS anisotropy (black). The 2D contours show 95\% CIs. The pink lines show the true parameters $\theta_{\rm mock}$.}
    \label{fig:corner}
\end{figure}

For computational convenience, we fix $E_0$ and $L_X$ to the true values in $\theta_{\rm mock}$ and perform the inference over the remaining three astrophysical parameters for which we assume flat priors over the following ranges: (i) $\log_{10} f_{\ast, 10} \in [-2, -0.5]$; (ii) $\log_{10}f_{\rm esc, 10} \in [-3, 0]$; and (iii) $\rm{M}_{\rm turn}[\rm{M}_\odot] \in [8,10]$.

We run the inferences with the \texttt{21cmMC} \footnote{\url{https://github.com/21cmfast/21CMMC}} (e.g., \citealt{Greig15, Greig17, Greig18}) package using the {\tt MultiNest} (e.g. \citealt{Feroz09}) \footnote{We choose {\tt MultiNest} over {\tt UltraNest} because the former requires significantly fewer likelihood evaluations than the latter and because we expect a simple ellipsoidal posterior.} sampler. Each inference requires about 15k likelihood evaluations. In the following section, we show and discuss the posterior from each of these inferences.

\subsection{Inference results}
In Figure \ref{fig:corner} we show the marginal posteriors of our three inferences, together with the true values, $\theta_{\rm mock}$ (marked in pink).  In orange, we show the posterior for inference (i).  As foreshadowed by the orange curves in Figure \ref{fig:hera_mock}, we see that the posterior is indeed biased, due to the mismatch of the 2D PS footprints of the observation and forward model.  The bias is at the level of $\sim10$\% in the inferred ionising escape fraction and stellar fraction parameters.  More dramatic however is the overconfidence of the biased posteriors, with the true values being outside $\sim 10\sigma$ for both $f_{\ast,10}$ and $f_{\rm esc, 10}$.
%being off by  our ionizing escape fraction parameters at the level of  when spherically-averging.  We can see that it is highly biased for all parameters, for a total bias of over $15\sigma$. In blue, we show the posterior for inference (ii) (analogous to the blue lines in Figure \ref{fig:hera_mock}). In black, we show the posterior for inference (iii) (analogous to the black dashed line in Figure \ref{fig:hera_mock}).

In blue, we show the posterior for inference (ii).  Again, as foreshadowed by the correspondingly-coloured curves in Figure \ref{fig:hera_mock}, this posterior is unbiased; however, it is notably wider than the posterior for inference (iii), shown in black.  This highlights the impact of sample variance.  
Applying \denoiser\ to a realisation of the 2D PS during inference tightens the inferred 1D marginal parameter constraints by $\sim50$\%.

\section{Conclusion} \label{sec:conclusion}

In this work, we study the consequences of two common approximations made on the 21-cm PS likelihood in Bayesian inference problems:
\begin{enumerate}[(i)]
    \item Replacing the 21-cm PS mean with a sample from a single realisation;
    \item Averaging the 3D Fourier modes within a $k$ bin over all orientations, instead of matching those actually observed by 21-cm PS experiments.
\end{enumerate}
In order to relax both of these assumptions, we developed \denoiser, a score-based diffusion model that provides an estimate of the mean cylindrical 21-cm PS given a single realisation. Unlike emulators, \denoiser\ is not tied to a particular model or simulator since its input is a (model-agnostic) realisation of the 2D 21-cm PS. \denoiser\ outperforms state-of-the-art analytical approaches such as fixing and pairing, the benchmark technique for sample variance mitigation. Individual 2D PS realisations can deviate from the mean by over 50\% at scales relevant to current interferometers. \denoiser\ reduces this deviation to $\sim$ 2\%, over a factor of 2 better than fixing and pairing, and at almost no additional cost ($\sim6$ s per iteration). Moreover, we test \denoiser\ on 21-cm power spectra from a completely different simulator and astrophysical model. We find that it produces reasonable power spectra and produces a mean estimate that is $\sim 2.5$ times more accurate than the 2D PS realisation itself above $\mu_{\rm min} = 0.97$, the region of cylindrical PS space most relevant to current observations. 

We test \denoiser\ by applying it in a realistic inference context. First, we simulate a realistic HERA mock 21-cm PS observation with {\tt 21cmSense}. Then, we run a set of three inferences: (i) a classical inference such as previous state-of-the-art inferences; (ii) an improved inference where we solve only the first issue mentioned above; and (iii) where we solve both issues. 
We find that inference (i), the typical state-of-the-art inference method, produces a highly overconfident and biased posterior with a bias of over $10\sigma$ for two of the three parameters. In inference (ii), we crop the $k$-space of our cylindrical 21-cm PS closer to that of observations by applying a cut at $\mu_{\rm min} = 0.97$, and, as expected, we obtain an unbiased but wider posterior. Finally, we apply both \denoiser\ and the cut at $\mu_{\rm min} = 0.97$ in inference (iii), and obtain an unbiased posterior that is on average 50\% narrower for each parameter.

We leave to future work to apply score-based diffusion models toward other sample-variance limited observations, such as large-scale structure surveys, and quantify the improvement they provide over previous machine learning methods such as convolutional neural networks (e.g., \citealt{DeSanti22}).

\section*{Acknowledgements}
D.B. thanks Laurence Perrault-Levasseur and Yashar Hezaveh for useful discussions and comments on an early draft of this project, as well as Alexandre Adam for assistance in implementing the score-based diffusion model. We thank Adrian Liu for useful discussions and comments on the manuscript.
We gratefully acknowledge computational resources of the Center for High Performance Computing (CHPC) at SNS.
A.M. acknowledges support from the Ministry of Universities and Research (MUR) through the PRIN project "Optimal inference from radio images of the epoch of reionization", and the PNRR project "Centro Nazionale di Ricerca in High Performance Computing, Big Data e Quantum Computing".
S. G. M. has received funding from the European Union’s Horizon 2020 research and innovation programme under the Marie Skłodowska-Curie grant agreement No. 101067043.
In addition to the packages references in the text, this work made use of the open-source Python packages {\tt NumPy} \citep{Harris20}, {\tt Matplotlib} \citep{Hunter07}, {\tt Astropy} \citep{astropy:2013, astropy:2018, astropy:2022}.
%%%%%%%%%%%%%%%%%%%%%%%%%%%%%%%%%%%%%%%%%%%%%%%%%%
\section*{Data Availability}

Both \denoiser\ and \texttt{tuesday} are on publicly accessible github repositories, as well as available for installation as a Python package using \texttt{pip}.

%%%%%%%%%%%%%%%%%%%% REFERENCES %%%%%%%%%%%%%%%%%%

\bibliographystyle{aa}
\bibliography{refs}

\end{document}